\documentclass[]{emulateapj}
\newcommand{\lirg}{L$_{\mathrm{IR}} \sim $10$^{11}$ L$_\odot$}
\newcommand{\rad}{$r_{200}$}
\shorttitle{}
\shortauthors{Webb et al.}

\begin{document}


\title{The Evolution of  Dusty Star formation in Galaxy Clusters to $z = 1$:  {\it Spitzer} IR Observations of  the First Red-Sequence  Cluster Survey}


\author{T.M.A. Webb\altaffilmark{1} }
%\affil{Astronomy Department, University of California,
  %  Berkeley, CA 94720}

%\author{C. D. Biemesderfer\altaffilmark{4,5}}
%\affil{National Optical Astronomy Observatories, Tucson, AZ 85719}
%\email{aastex-help@aas.org}

%\and

\author{D. O'Donnell\altaffilmark{1}}
\author{H.K.C. Yee\altaffilmark{2}}
\author{David Gilbank\altaffilmark{4}}
\author{Kristen Coppin\altaffilmark{1}}
\author{Erica Ellingson\altaffilmark{3}}
\author{Ashley Faloon\altaffilmark{1}}
\author{James E. Geach\altaffilmark{1}}
\author{Mike Gladders\altaffilmark{5}}
\author{Allison Noble\altaffilmark{1}}
\author{Adam Muzzin\altaffilmark{6}}
\author{Gillian Wilson\altaffilmark{7}}
\author{Renbin Yan\altaffilmark{8}}

%\affil{Space Telescope Science Institute, Baltimore, MD 21218}


\altaffiltext{1}{McGill University, 3600 rue University, Montreal, QC, Canada, H3A 2T8}
\altaffiltext{2}{Department of Astronomy and Astrophysics, University of Toronto, 50 St. George St., Toronto, ON, Canada, M5S 3H4 }
\altaffiltext{3}{Department of Astrophysical and Planetary Sciences, University of Colorado at Boulder, Boulder, CO, USA, 80309}
\altaffiltext{4}{South African Astronomical Observatory, PO Box 9, Observatory, 7935, South Africa}
\altaffiltext{5}{Department of Astronomy and Astrophysics, University of Chicago, 5640 S. Ellis Ave., Chicago, IL, USA, 60637}
\altaffiltext{6}{Leiden Observatory, University of Leiden, Niels Bohrweg 2, NL-2333 CA, Leiden,  The Netherlands}
\altaffiltext{7}{Department of Physics and Astronomy, University of California at Riverside, 900 University Avenue, Riverside, CA, USA, 92521}
\altaffiltext{8}{Center for Cosmology and Particle Physics, Department of Physics, New York University, 4 Washington Place, New York, NY, USA, 10003}
%\altaffiltext{2}{Society of Fellows, Harvard University.}
%\altaffiltext{3}{present address: Center for Astrophysics,
%    60 Garden Street, Cambridge, MA 02138}
%\altaffiltext{4}{Visiting Programmer, Space Telescope Science Institute}
%\altaffiltext{5}{Patron, Alonso's Bar and Grill}


\begin{abstract}
We present the results of an infrared (IR) study of high-redshift  galaxy clusters with the MIPS camera on board the {\it Spitzer Space Telescope}.
We have assembled a sample of 42 clusters from the Red-Sequence Cluster Survey-1  over the redshift range 0.3 $< z < $ 1.0 and
spanning an approximate range in mass of 10$^{14-15}$M$_\odot$.   We statistically measure the number of IR-luminous galaxies in clusters above a fixed inferred IR luminosity of 2$\times$10$^{11}$ M$_\odot$, assuming a star forming galaxy template,  per unit cluster mass  and find it increases to higher redshift.  Fitting a simple power-law we measure evolution of $(1+z)^{5.1\pm1.9}$ over the range  0.3 $< z < $ 1.0.   These results are  tied to the adoption of a single star forming galaxy template; the presence of AGN, and an evolution in their relative contribution to the mid-IR galaxy emission, will alter the overall number counts per cluster and their rate of evolution.  Under the star formation assumption we infer the approximate total SFR per unit cluster mass ($\Sigma$SFR/M$_\mathrm{cluster}$).  
The evolution is similar, with $\Sigma$SFR/M$_\mathrm{cluster} \sim (1+z)^{5.4\pm1.9}$. We show that this can be accounted for by the evolution of the IR-bright field population over
the same redshift range; that is,  the evolution can be attributed entirely to the change in the in-falling field galaxy population. We show that the $\Sigma$SFR/M$_\mathrm{cluster}$ (binned over all redshift) decreases with increasing cluster mass with a slope  ($\Sigma$SFR/M$_\mathrm{cluster} \sim$ M$_\mathrm{cluster}^{-1.5\pm0.4}$) consistent with the dependence of
the stellar-to-total mass   per unit cluster mass seen locally. The inferred star formation seen here could produce $\sim$5-10\% of the total stellar mass in massive clusters at $z = 0$, but we cannot constrain the descendant  population, nor how rapidly the star-formation must shut-down once the galaxies have entered the cluster environment.  Finally, we show a clear decrease in the number of IR-bright galaxies per unit optical galaxy in the cluster cores, confirming star formation continues to avoid the highest density regions of the universe at $z \sim$ 0.75 (the average redshift of the high-redshift clusters).  While several previous studies appear to show enhanced star formation in high-redshift clusters relative to the field we note that these papers have not accounted for the  overall increase in galaxy or dark matter density at the location of clusters.  Once this is done, clusters at $z\sim$ 0.75 have the same or less star formation per unit mass or galaxy as the field.

\end{abstract}





\section{Introduction}

%SFR density  relation - local and high redshift.  (Hashimoto 1998, Lewis 2002, Gomez 2003)  

%Butcher Oemler effect (Butcher Oemler 1984)

%Quenching vs downsizing etc etc

%cluster SFR/mass Kodama 04 Finn 04/05 Geach 06 Bai 09 Chung 10 Koyama 10 Hayashi 11

%ref also Saintonge and Haines -- but these are fractions rather than sum.

%SFR density relation:  environmental effect or advanced state of evolution in clusters relative to the field (nature vs nurture). 
%need for large samples, systematic selection, wide redshift range - out to z = 1

%compare different selection techniques?  Haines claims BO effect is reduced in X-ray samples (refs = Andreon 2006, Fairley 2002, Ellingson 2001) compared to optical  (De Propis 2007, Goto 2007) 

%see debate on cluster properties vs fb (Margoniner 2001 Wake 2005 Goto 2003 Popesso 2007 Aguerri 2007 
%claim is that for sigma > 600 km/s X-ray there is no evolution in fb to z = 0.8  also Nakata 05 and Andreon 2006

%go over other IR BO studies - Haines and Saintonge

It is now clear that the environment is a primary factor in  galaxy evolution, either through direct influences, or because galaxy density is a tracer of the important underlying drivers of evolution, such as galaxy mass or formation time \citep[e.g.,][]{gomez03,baldry06}.  Moreover, the importance of the environment may be a strong function of cosmic epoch; even if  environmental dependencies are constant with redshift \citep{peng10}, galaxies themselves are located in progressively more dense regions with time.   On the other hand, properties of galaxies such as  their mass are also  predictors of their evolution   and  trends with density are in part driven by underlying mass-density biases. \citep{peng10,muzzin12}.    Understanding the complex interdependencies of mass and environment over the history of the universe, and their effect on galaxy formation,  is an immense observational endeavor, requiring substantial dynamic range in galaxy and halo mass,  environment or density,  and time. 

At $z = 0$, star formation is suppressed in high-density regions \citep[e.g.,][]{kauffmann04}, but recent studies have shown that by $z \sim$ 1 star formation has begun to migrate from low to higher-density environments \citep{elbaz07,cooper08}. Still,  the highest density regions at any given epoch - galaxy clusters - are difficult to probe precisely because these regions are so rare.  While wide-field surveys probe several orders of magnitude in galaxy density, most do not contain large numbers of galaxy clusters, and for these densities targeted investigations are more efficient.
However, the assessment of the star formation rates of galaxies in high-redshift clusters has been hindered in the past by the inhomogeneity and sparseness of cluster samples, and the different approaches used to quantify the amount of star formation (SF) occurring in the cluster environment.  

Conclusions drawn from the comparison of small numbers of clusters selected through different biases are problematic.   
A number of lines of evidence suggest that the star formation efficiency of a cluster is strongly correlated with its total halo mass (though not necessarily driven by it) \citep[e.g.,][]{poggianti06}. The dynamical state of a cluster at the epoch of observation is likely also very important, as a major merger event could have a profound though temporary effect on the member galaxies.  
Comparisons between different studies are further complicated by star formation rates that are estimated through different observational diagnostics such as optical emission lines ([OII], H$\alpha$) or infrared emission.  These different techniques are  biased to different galaxy populations, or may yield  different values for the same galaxy either through errors in the calibrations or by probing different star-forming regions. Finally, studies have employed a variety of different cluster member identification methodologies: spectroscopic confirmation, photometric redshifts or color selection, each with different observational limits and completeness functions.  These different approaches have all shown some level of evolution in the SFR of clusters to higher redshift but have varied significantly in their assessment of its magnitude: $f_{SF} \propto (1+z)^{2-7}$ \citep{kodama04,geach06,saintonge08,bai09,haines09,popesso12}

Here we present a study of the average IR properties of a large sample of galaxy clusters (42)  drawn in a systematic way from the first Red-Sequence Cluster Survey \citep{rcs05} over  the redshift range 0.3 $ < z < $ 1.0.  The intent of this work is to undertake a very simple analysis of the global evolution of dust enshrouded activity in cluster environments with time.  Our adopted method offers a number of advantages, but is also limited in scope and for clarity we briefly summarize it here.  We perform a simple statistical measurement of the  background subtracted IR counts along the line of sight to galaxy clusters.  This provides a clean measurement of the dust-enshrouded activity in clusters that does not require any selection or assumptions beyond a simple IR-luminosity cut.    We do not, for example, require an optical counterpart for the IR sources, which would bias the sample toward less dust-enshrouded objects; nor do we  directly isolate cluster members, which then requires a member completeness correction and is again biased.  Thus, we will be sensitive to {\it all} activity  above our detection limit. Throughout we make the simplification that the IR luminosity is entirely produced by star-formation,  with no AGN contamination.   This assumption is likely an oversimplification \citep{tomczak2011} and does not affect the primary conclusions of the paper which depend only on the IR luminosity, not the source of activity.   We discuss AGN throughout the paper, when they are relevant to the conclusions.

On the other hand, this  statistical method limits us to the cluster averaged activity and cannot tell us anything about the individual galaxies.   Therefore we cannot control for galaxy specific properties such as stellar mass and this limits the implication of the results.
 More detailed analysis of the  IR galaxy population in this cluster sample will be undertaken in future papers, nevertheless, important conclusions may be drawn from this simple study alone.

 The paper is laid-out as follows.  
In \S 2 we outline the RCS cluster sample selection and  describe the {\it Spitzer} observations, data reduction and source extraction. In \S 3 we describe the number count analysis methodology and in  \S 4 we present the results.  \S 4.1 presents the evolution of the statistical excess of IR galaxies seen in the RCS cluster fields and  describes sources of systematic error in this analysis; \S 4.2 converts these measurements to the evolution of the integrated star formation per unit cluster mass;  \S 4.3 compares this evolution to that of the IR-bright field population; in \S 4.4 we look at the dependence of these results on cluster richness or mass; and \S 4.5 presents the radial distribution of the IR-luminous population in clusters.  In \S 5 we discuss the implications of these results and in \S 6 we summarize the primary conclusions of the paper. We use H$_\circ$ = 70 km/s/Mpc, $\Omega_\mathrm{M}$ = 0.3 and $\Omega_\Lambda$ = 0.7 throughout. 

%Describe cosmology - at the moment this is not consistent:  we use a different cosmology than the RCS catalog and Yee \& Ellingson's relations.

 %     Galaxies may also undergo episodes of enhanced star formation during their accretion, triggered by various means such as merging or harassment  (REFS).  

\section{Observations and Data Reduction}

\subsection{The Red-Sequence Cluster Survey Sample}

The cluster sample was drawn from the first Red-Sequence Cluster Survey (RCS-1).  The RCS-1 is a 90 square degree optical imaging survey conducted in two filters ($R_c$ and $z^\prime$) with the CTIO and CFHT telescopes \citep{rcs00}. It was designed to optimize the detection of galaxy clusters in the redshift range $0.3 < z < 1.2 $ through the detection of  the red sequence of early-type galaxies within the cluster core.  The cluster finding method has proven to be extremely robust and suffers from minimal projection effects; X-ray imaging, spectroscopic verification and simulations consistently show this to be $<$ 10\% \citep{hicks08,gilbank07}.  Moreover, the localization of the red-sequence in color space constrains the cluster redshift to $\sim$5\%.  The two-band photometry further provides an estimate of the richness through the parameter N$_\mathrm{red}$ \citep{lu09}, the number of red-sequence galaxies brighter than M$^\star$+ 2 within 0.5Mpc.  Throughout the paper we use the RCS-1 estimated N$_\mathrm{red}$ which uses the  $z^\prime$ magnitude and ($R_c$ - $z^\prime$) color to define red sequence galaxies. We note that this is a version of the cluster catalog updated from previous publications by our group, and will be presented in Barrientos et al. (in preparation).  The richness measurements will be discussed in detail in Ellingson et al. (in preparation) and Gilbank et al. (in preparation)\footnote[1]{We note that this richness measurement is essentially the same as the Bgc parameter we have used in previous work but uses a somewhat larger counting radius and does not extrapolate the luminosity function beyond the observational limits. Our previous Bgc measurements may be approximately converted to N$_{red}$ by dividing the former by $\sim$30}. 

% is a re-analysis of the RCS-1 data and cluster catalog that will be discussed in detail in Ellingson et al. (in preparation). 

\begin{figure}
  \includegraphics[scale=0.5]{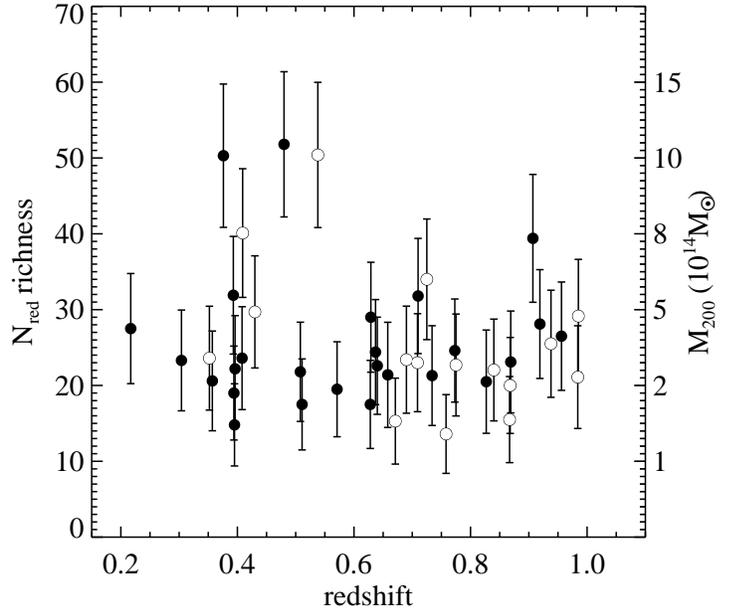}
%\vskip 0.7cm
\caption{The optical richness, parametrized by  N$_\mathrm{red}$ \citep{lu09} versus the redshift of each cluster in the RCS-MIPS sample presented here. Solid points denote clusters with spectroscopic redshift confirmation and open points correspond to redshifts determined through a fit to the color location of the red sequence.  The corresponding M$_{200}$ was computed following the N$_\mathrm{red}$-$\sigma$ relation of Ellingson et al. (in preparation).  The uncertainties can be applied to both axes, but in the case of M$_{200}$ do not include any intrinsic scatter in the richness-mass relation. \label{bgcz}}
\vskip 0.5cm
\end{figure}

From the RCS-1 parent sample, we selected 42 clusters to uniformly fill  the redshift range 0.3 $< z < $ 1.0 and to span a richness range which corresponds to roughly an order of magnitude in mass (N$_\mathrm{red}=$10-60, and using the N$_\mathrm{red}$ to mass conversion outlined below,  M $ \sim$ 10$^{14-15}$ M$_\odot$, with $\sim$ 30\% uncertainties).   In so doing,  we have assembled a sample which is  no longer  representative  of the RCS selection distribution in mass or redshift, but rather  comprises a representative sample with the goal of isolating  the effects of redshift and cluster properties on cluster galaxy evolution.  The size of the sample was motivated by the desire to remove uncertainties due to cluster-to-cluster variations, by averaging over bins of redshift and/or richness with a minimum of ten clusters per bin.   Figure \ref{bgcz} shows the richness and redshift values for the cluster sample; no selection bias in richness with redshift is evident.  Table \ref{data_table} summarizes relevant cluster properties.

The RCS-1 clusters are the focus of a number of spectroscopic campaigns by the RCS consortium  for the purpose of redshift confirmation, population and cluster dynamic studies, and gravitational lensing analyses, which have provided extensive but inhomogeneous coverage of the RCS-MIPS fields. In Figure \ref{bgcz} and Table \ref{data_table}  we indicate the 25/42  clusters with spectroscopic redshift confirmation.  The larger spectroscopic sample have also calibrated the relation between the N$_\mathrm{red}$ richness and the velocity dispersion and this is presented in Ellingson et al. (in preparation).  

 In this paper we work within a radial limit of $r_{200}$ of each cluster and also normalize by the  cluster M$_\mathrm{200}$ mass. The effects  the uncertainties on these properties have on our results are discussed in detail in the Appendix. 
We calculate the $r_{200}$ radii by first estimating the velocity dispersion from the optical richness as outlined above.  The $r_{200}$ radii were  determined through the following relation \citep{carlberg96}:

%The employed relation is shown in Figure 2 along with reliable velocity dispersion measurements for 10 RCS-MIPS clusters; note however that the relation was calibrated through a larger set of 50 RCS-1 clusters with IMACS spectroscopy and 20-100 members each.   

\begin{equation}
r_{200} = {\sqrt{3}\sigma \over{10 \mathrm{ H(z)}}}
\end{equation}

The  mass can then be calculated by:

\begin{equation}
M_{200} = {3\sigma^2 r_{200} \over{G}}
\end{equation}

\begin{deluxetable*}{cccccc}
\tablewidth{0pt}
\tablecaption{Information on cluster sample and Spitzer-MIPS observations \label{data_table}}
\tablehead{
\colhead{Cluster ID}             & \colhead{redshift\tablenotemark{a}}  &
\colhead{N$_\mathrm{red}$}          & \colhead{$r_{200}$}  & \colhead{integration} & \colhead{map size}\\
\colhead{} &   \colhead{} & \colhead{} & \colhead{(arcminutes)} & \colhead{time (s)} & \colhead{(arcminutes$^2$)} }
\startdata
RCS212134$-$6335.8  & \bf{0.217} &  27.5 $\pm$ 7.3 & 7.6 $\pm$ 1.8 & 200 & 225  \\
RCS035139$-$0956.4 & \bf{0.304} &  23.3 $\pm$ 6.6 & 5.2  $\pm$ 1.3 &200 & 225  \\
RCS022516+0011.5   & \bf{0.357} & 20.6  $\pm$ 6.6 & 4.3 $\pm$ 1.2 & 200 & 225   \\
RCS132655+3021.1  & 0.35 & 23.6 $\pm$ 6.8 & 4.6 $\pm$ 1.2 & 200 & 300  \\
RCS144726+0828.3  & \bf{0.376} & 50.3 $\pm$ 9.4 & 6.5  $\pm$ 1.1 & 200 & 300 \\

RCS092821+3646.5  & \bf{0.393} & 27.0 $\pm$ 7.8 & 4.9 $\pm$ 1.3 & 200   & 300 \\
RCS022359+0126.1  & \bf{0.394} &19.0 $\pm$ 6.2 & 3.7  $\pm$ 1.1 & 200 & 225 \\

RCS145226+0834.6  & \bf{0.395} &  14.8 $\pm$  5.4 & 3.3  $\pm$ 1.1  & 200 & 225 \\
RCS051834$-$4325.1  & \bf{0.396} & 21.8 $\pm$ 7.0 & 4.1 $\pm$ 1.2 & 200 & 300 \\

RCS022403$-$0227.7  & \bf{0.408} & 23.6 $\pm$ 6.8 & 4.1 $\pm$ 1.1 & 200 & 225 \\
RCS231526$-$0046.7  & 0.40 & 40.1 $\pm$ 8.5 & 5.4  $\pm$ 1.0 & 200 & 300 \\

RCS044207$-$2815.0  & 0.45 & 29.7 $\pm$ 7.4 & 4.4  $\pm$ 1.0 & 200 & 100 \\
RCS215223$-$0503.8  & \bf{0.480} & 51.8 $\pm$ 9.6 & 5.4  $\pm$ 0.9 & 200 & 150 \\

%RCS222717$-$5956.6  & 0.658 &  \hspace{4pt}658.9 $\pm$ 245.8 & 2.5 & & 10$\times$10\\
RCS051855$-$4315.0  & \bf{0.508} &  21.8 $\pm$ 6.5 & 3.3 $\pm$ 0.9 & 200  & 100  \\
RCS110733$-$0520.6  & \bf{0.511} &  17.5 $\pm$ 6.0 & 2.9 $\pm$ 0.9 & 600 & 100 \\
RCS234717$-$3634.4  & 0.55 & 50.4 $\pm$ 2.6 & 9.6 $\pm$ 0.4 & 200 & 150 \\
RCS110104$-$0351.3  & \bf{0.571} &  19.5 $\pm$ 6.3 & 2.7 $\pm$ 0.8 & 600 & 150 \\
RCS144654+0827.0 & \bf{0.628} & 17.5 $\pm$ 5.8 & 2.4 $\pm$ 0.7 & 200 & 100 \\
RCS144557+0840.3  & \bf{0.629} & 29.0 $\pm$ 7.2 & 3.2 $\pm$ 0.7 &  200& 150 \\
RCS110439$-$0445.0  & \bf{0.637} & 24.4 $\pm$ 6.9 & 2.9 $\pm$ 0.7  & 600 & 100  \\
RCS215248$-$0609.4  & \bf{0.649} &  22.6 $\pm$ 6.4 & 2.7 $\pm$ 0.7  & 3300 & 100  \\
RCS211852$-$6334.6  & \bf{0.658} &  21.4 $\pm$ 6.9 & 2.6 $\pm$ 0.7 & 200& 100 \\
RCS110246$-$0426.9  & 0.70 &  15.3 $\pm$ 5.7 & 2.1 $\pm$ 0.7 & 600  & 200 \\
RCS212238$-$6146.1  & 0.70 & 23.4 $\pm$ 7.1 & 2.6 $\pm$ 0.7 & 200 &100 \\
RCS112225+2422.9  & 0.70 & 23.0 $\pm$ 6.5 & 2.5 $\pm$ 0.6 & 2000  & 100 \\
RCS141910+5326.1 & \bf{0.710} & 31.8 $\pm$ 7.6 & 2.5 $\pm$ 0.4 & 2000 & 100 \\
RCS234220$-$3534.3  & 0.70 &  34.0  $\pm$ 8.0 & 3.0  $\pm$ 0.6 & 200 &100 \\
RCS044126$-$2813.2  & \bf{0.734} &  21.3  $\pm$ 6.6 & 2.4 $\pm$ 0.7 & 100 & 100 \\
RCS132939+2853.3  & 0.75 &  13.6 $\pm$ 2.0 & 5.2 $\pm$ 0.7 & 200 & 50 \\
RCS022433$-$0002.3  & \bf{0.773} &  24.6 $\pm$ 6.8 & 2.0 $\pm$ 0.5  & 3300 & 100   \\
RCS110411$-$0337.5  & 0.80 & 22.7 $\pm$ 6.7 & 2.4 $\pm$ 0.6 & 600 & 100 \\
RCS051940$-$4402.1  & \bf{0.827} &  20.5 $\pm$ 6.8 & 2.1 $\pm$ 0.6 & 200 & 50 \\
RCS110118$-$0328.6  & 0.80 &  22.0 $\pm$ 6.7 & 2.2 $\pm$ 0.6 & 600 & 50 \\
RCS110206$-$0414.5  & 0.90 &  15.5 $\pm$ 5.7 & 1.8 $\pm$ 0.6 & 600 & 50  \\
RCS110615$-$0330.8  & 0.90 &  20.0 $\pm$ 6.3 & 2.0  $\pm$ 0.6 & 600 & 100 \\
RCS162009+2929.4  & \bf{0.869} &  23.1 $\pm$ 6.7 & 2.2 $\pm$ 0.6 & 2000  & 100 \\
RCS231953+0038.0  & \bf{0.907} & 39.4 $\pm$ 8.4 & 2.8  $\pm$ 0.5 & 3200 & 100 \\
RCS132631+2903.3 & \bf{0.919} & 28.1 $\pm$  7.2 & 2.3  $\pm$ 0.5 & 200 & 100 \\
RCS022158$-$0340.1  & 0.90 &  25.5 $\pm$ 7.1 & 2.1 $\pm$ 0.5 & 4750 & 50 \\
RCS051908$-$4323.3  & 1.0 & 21.1 $\pm$ 6.8 & 1.8 $\pm$ 0.5 & 200 & 100 \\
RCS022056$-$0333.3  & 1.0 & 29.2 $\pm$ 7.5 & 2.2 $\pm$ 0.5 & 4750 & 50 \\
RCS043938$-$2904.8  & \bf{0.956} & 26.5 $\pm$ 7.1 & 2.2 $\pm$ 0.5 & 2000 & 100 \\
%RCS132629+2903.1  & 1.046 &  1901.6 $\pm$ 451.2 & 3.2 & & 10$\times$10 \\
\enddata
\tablenotetext{a}{Redshifts in bold are spectroscopically confirmed; non-bold values denote
redshifts determined through a fit to the location of the cluster red-sequence, good to $\sim$5\%.} 
%make this footnote extra long so that it extends over two lines.}
%% You can append references to a table using the \tablerefs command.
%\tablerefs{
%(1) Barbuy, Spite, \& Spite 1985; (2) Bond 1980; (3) Carbon et al. 1987;}
\end{deluxetable*}

%\begin{figure}
%\epsscale{1.2}
%\plotone{plot_veldisp.ps}

%\hskip 1cm \includegraphics[scale=0.9]{plot_veldisp.ps}
%\vskip 1cm
%\caption{ {\it Top:} A comparison of the N$_\mathrm{gal}$ richness estimator with the measured velocity dispersion  of the cluster. The solid line shows the best fit relationship using the larger RCS-1 spectroscopic catalog ($\sim$50 clusters; Ellingson et al., in preparation),  used here to determine M$_{200}$ and $r_{200}$.   {\it Bottom:} A comparison of the redshift determined through red-sequence fitting and the confirmed spectroscopic redshift of the cluster, for the 26/42 clusters with spectroscopic confirmation; here we require 5 confirmed members within 1Mpc. The location of one-to-one correspondence is shown. \label{catalog} }
%\end{figure}
 
\subsection{{\it Spitzer} Imaging: MIPS 24$\mu$m and IRAC 3.6/4.5$\mu$m}

Our primary data set, the Spitzer 24$\mu$m imaging, was obtained through open time program 30940. The MIPS observations were designed to reach an approximate depth of {\lirg} out to a radius of {\rad} for each cluster  and therefore per pixel integration times and image sizes vary  depending on cluster richness, redshift and the thermal background. A subset of the clusters obtained longer integrations to facilitate a deeper study on a smaller number of clusters,  but this additional depth is not relevant for this work.   The  integration time and map area for each cluster are listed in Table \ref{data_table}; average exposure times per pixel  range from 200s to 4750s and image areas from 50 arcmin$^2$ to 300 arcmin$^2$, totaling $\sim$1.5 deg$^2$.  The MIPS images were reduced using a combination of the Spitzer Science Center's MOPEX software and our own IDL routines developed to further optimize background subtraction in each field. 
%The in 24$\mu$m background is dominated by zodiacal light, with additional contributions from interstellar cirrus emission and the cosmic background of unresolved galaxies. The distribution of zodiacal light and interstellar cirrus vary appreciably with ecliptic and galactic latitude respectively and make consistently effective reduction challenging when working with targets spread across the sky, as this sample is, and variations in the map depths  (beyond the usual $sqrt{t}$ scaling) is expected. 

%
In  \S 4.4 and  the Appendix we will compare the MIPS measurements to a comparison IRAC-selected (rest-frame NIR) population.  For this we use deep IRAC 4-channel imaging which was obtained through open time program ID 20754 for all $z > $ 0.5 clusters within the RCS-MIPS sample, with the exception of RCS022056 and RCS022158. These two clusters are located within the public SWIRE fields and we do not use these two clusters in the IRAC-dependent analysis to avoid introducing systematic errors due to the different observational mode of the SWIRE data. The goal of the program was to probe the stellar mass of the cluster galaxies and therefore the IRAC sample was limited to $z >$ 0.5;  at lower redshifts NIR imaging samples the appropriate rest-frame wavelength and these clusters are part of an on-going CFHT imaging program.  The IRAC and CFHT observations were designed to reach an approximate depth of M$^\star$+1.5 at each redshift. 

Clusters with redshifts  below $z\sim$ 0.78 were centred on the 3.6$\mu$m array and those above this redshift were placed on   the 4.5$\mu$m array;  this ensured sampling at the observed wavelength which most closely corresponded to rest-frame K.  Image processing was performed using {\it IRACproc} \citep{schust06} a software suite developed to wrap the existing MOPEX pipeline in IRAC mode and add IRAC specific improvements to the outlier rejection. 

\subsection{Source Detection and Photometry Catalog}

The 24$\mu$m source extraction algorithm combined the source-finding capabilities of the PPP detection and photometry program \citep{yee91} with  the aperture photometry and PSF-fitting capabilities of DAOPHOT \citep{stet87}.  Source detection and aperture photometry on the IRAC images was done using PPP \citep{yee91} alone.    

\begin{figure}
\hskip -1cm \includegraphics[scale=0.5]{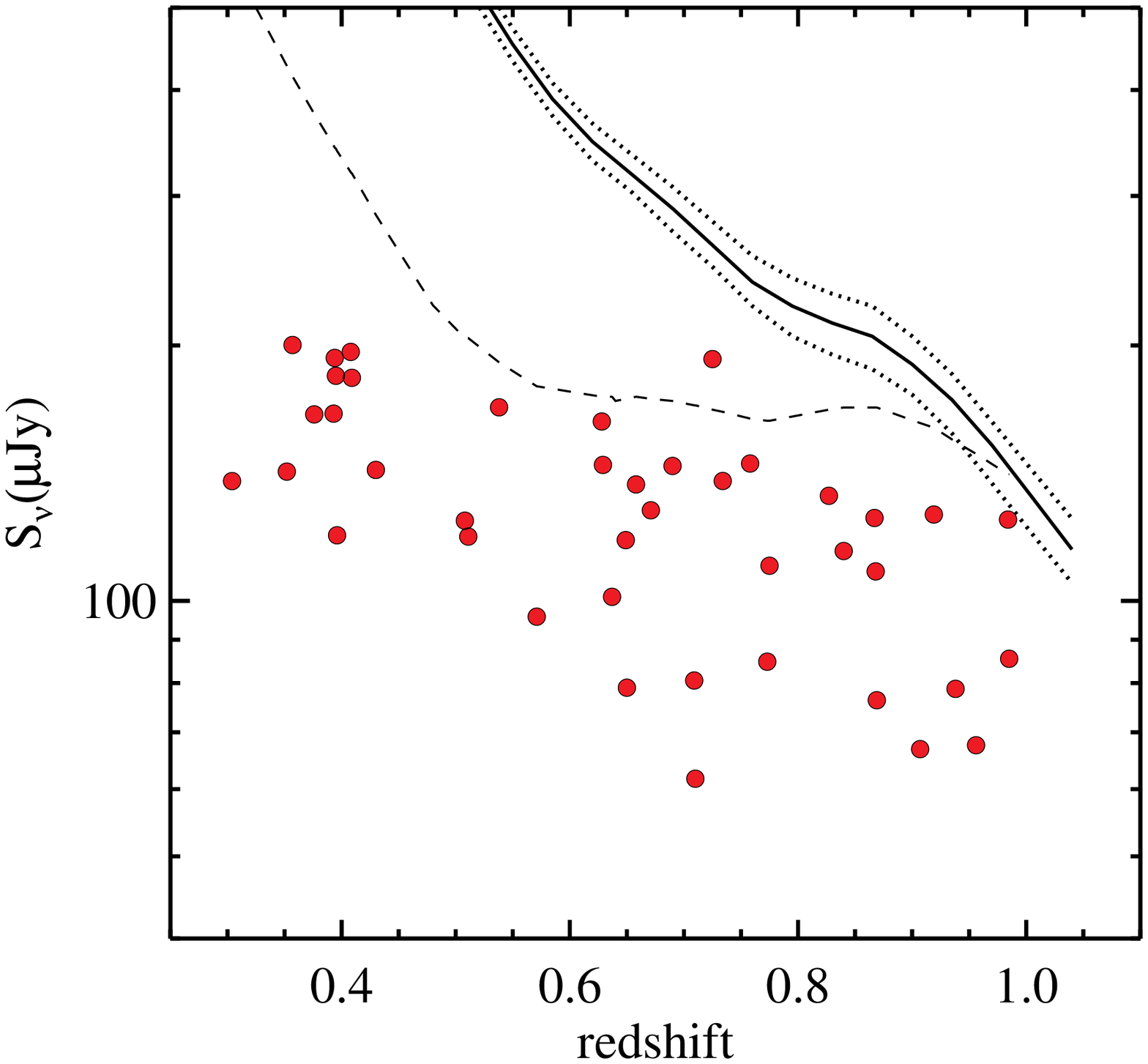}
\caption{An illustration of the MIPS image depth for each cluster field  compared to the expected 24$\mu$m flux of a template galaxy with L$_{\mathrm{IR}} $ = 2$\times$10$^{11} $L$_\odot$; the solid line corresponds to the combined models of \citet{ce01} and \citet{dh02} and the two dotted lines show the range in these two model libraries.  The solid points show the $\sim$80\% completeness level for each separate cluster field as determined by the Monte Carlo simulations described in the text.
The dashed line denotes the L$_\mathrm{IR}^\star$ parametrized depth used in \S 4.2: we set the $z = $ 1 depth to    L$_\mathrm{IR}$ =  2$\times$10$^{11} $L$_\odot$ and alter the $z < $ 1 depths to follow the evolution of    L$_\mathrm{IR}^\star$ of the field.   In \S 4.1 we count all galaxies above the solid line and in \S 4.2 we count above the dashed line. 
  \label{depth} }
 \vskip 0.5cm
\end{figure}

Completeness limits  at 24$\mu$m were estimated through the insertion and recovery of fake sources into each cluster image, using the same detection and photometry method as used for the un-altered images. An input source was deemed recovered if an object was found within 2$\arcsec$ of the input position.  No secondary check against the true source catalog was made, and no input/output flux ratio criterion was employed. While this may lead to a slight underestimate of the actual completion limits it  requires the fewest assumptions and subjective definitions of a recovered object.   Figure \ref{depth} shows the 80\% completeness limit for each cluster field, compared with the expected 24$\mu$m flux of an infrared-luminous galaxy with L$_\mathrm{IR} \sim$  2$\times$10$^{11}$ L$_\odot$ over the same redshift range \citep{ce01,dh02}.

 Fields with two MIPS AORs taken at different times reveal a small number of asteroids.  Theoretical predictions of the Tedesco's Statistical Asteroid Model \citep{tedesco05} and the results of other extragalactic programs \citep{papovich04} indicate that the number of asteroids per field as well as the spread in number density between the ecliptic and zenith does not significantly alter the galaxy counts.   We therefore do not attempt to identify or remove the asteroid contribution in the cluster  fields with multiple AORs.

 %Monte Carlo simulations as follows. We inserted five artificial PSFs into each cluster image over twenty separate realizations. This was repeated for twenty flux bins between 50 and 330 $\mu$Jy. Each modified image was then passed through the same detection and photometry pipeline as the real data. {\it This needs some consideration and explanation.   Raw completeness is simply whether or not a source is recovered at any flux.  Should also have the info to look at recovered flux.  No bias here until the fields become incomplete: below the turnover in completeness there are fewer sources per flux bin than input, although all of those sources are in fact recovered at higher levels presumably.  Should probably include a table here with exp times and depths.}

\subsection{Background Comparison Field: SCOSMOS}

\begin{figure}
\hskip -1cm \includegraphics[scale=0.5]{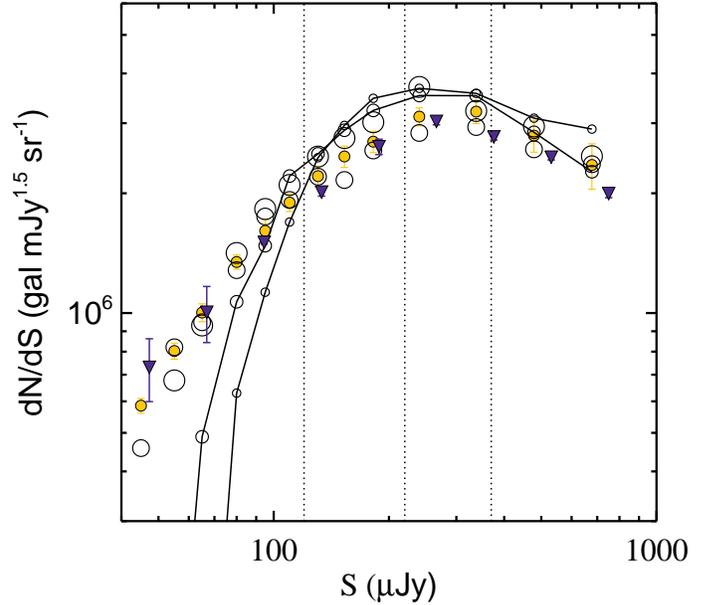}
\caption{The Euclidian-normalized differential number counts of MIPS sources in the cluster fields compared to \citep{papovich04} (blue diamonds) and our analysis of the SCOSMOS GO3 data (solid orange circles). The uncertainties on the Papovich counts do not include field-to-field variance. The open circles correspond to the cluster fields divided into four integration time bins:  200s, 600s, 2000s, $>$ 3200s.  The size of the circle increases with integration time. The two solid lines connect the counts for the 200s and 600s images, to draw the eye to the region where they begin to diverge from the SCOSMOS counts due to incompleteness, at $\sim$100$\mu$Jy. The longer integration time images are consistent with COSMOS to lower levels.   Also shown (dotted lines) are the inferred luminosities of a galaxy with L$_{\mathrm{IR}}$ = 2$\times$10$^{11}$L$_\odot$ (LIRG) for $z = $ 0.6, 0.8 and 1.0 that we employ in this work ($z= $ 0.4 lies off the plot).   We count galaxies to the right of the dotted lines for each redshift bin, and thus this illustrates the sufficient depth of all of the images, given the adopted limit,  and the adequate depth of the  comparison SCOSMOS field.  \label{diff_counts}}
\vskip 0.5cm
\end{figure}

The analysis presented here requires measurements of the background field galaxy population (i.e. along the line of sight). We use the publicly available SCOSMOS GO3 24$\mu$m data \citep{sand07} which is of comparable depth as our deepest cluster images and covers 2 deg$^2$ (Figure \ref{diff_counts}).  
To avoid introducing systematic errors into our analysis we worked directly with the original SCOSMOS unreduced data (rather than the publicly available source catalogs or images) and reduced/mosaicked them in an identical manner as the cluster fields. 

The reduced SCOSMOS comparison field was then passed through the same detection and photometry pipeline as the clusters.  In Figure \ref{diff_counts} we show the resulting differential number counts (Euclidian-normalized) of  the cluster fields  and our SCOSMOS catalogs and compare with the published counts of \citet{papovich04}.  Good agreement is seen between our SCOSMOS analysis and Papovich et al., while the clusters show a slight over-all excess which will be discussed throughout the paper.  We have also averaged the cluster counts over four exposure time bins to again illustrate the difference in depth within the cluster sample. The flux at which the cluster counts diverge from the field counts due to completeness (i.e. where the cluster count curve falls below the field)  for each bin agrees well with the MC results described above.

\section{Analysis: Galaxy Counts and Background Subtraction}

%\subsection{Galaxy Counts and Background Subtraction}

To estimate the observed 24$\mu$m flux limit at each cluster redshift, corresponding to a constant luminosity limit, we employ the models and prescriptions of \citet{ce01} and \citet{dh02}.  Implicit in this is the assumption that the shape of the IR SED of galaxies does not evolve appreciably over the redshift range of interest.   We work to the limit of L$_\mathrm{IR}$ $\sim$2$\times$10$^{11} $ L$_\odot$: this is set by the completeness limits of the cluster MIPS images (Figure \ref{depth})  and indeed the exact luminosity limit is less important than its consistency across all cluster fields.  Following \citet{bell03} this limit corresponds to a star formation rate of $\sim$30 M$_\odot$yr$^{-1}$; although rare in the local universe this is a more representative level of field galaxy star formation by $z>$ 0.5 \citep{lefloch05}.   In this analysis we have assumed the contamination of the MIR emission from AGN is negligible.   In doing so we may be introducing systematic uncertainties into the measurements and this is discussed in more detail in the Appendix.

Background counts are determined in a similar manner.  For each cluster we measure the average number of galaxies above its calculated flux limit and within the corresponding radius in 500 randomly placed apertures in the COSMOS field.  The standard deviation of these counts provide us with an estimate of the field-to-field variance in the background, and this is what we take to be the dominant uncertainty on each cluster measurement.   We do not attempt to incorporate the error in $r_{200}$ stemming from the richness uncertainties,  the intrinsic scatter in the mass-richness relation, any random error in the redshift, nor any systematic error in the flux-limit estimate.  Random uncertainties will be minimized by our stacking technique and systematics only matter if they are redshift dependent. We discuss the sources of error in detail in Appendix A. 

\section{Results}

\subsection{The number counts of IR-Luminous galaxies in cluster fields}

\begin{figure*}
\includegraphics[scale=0.6]{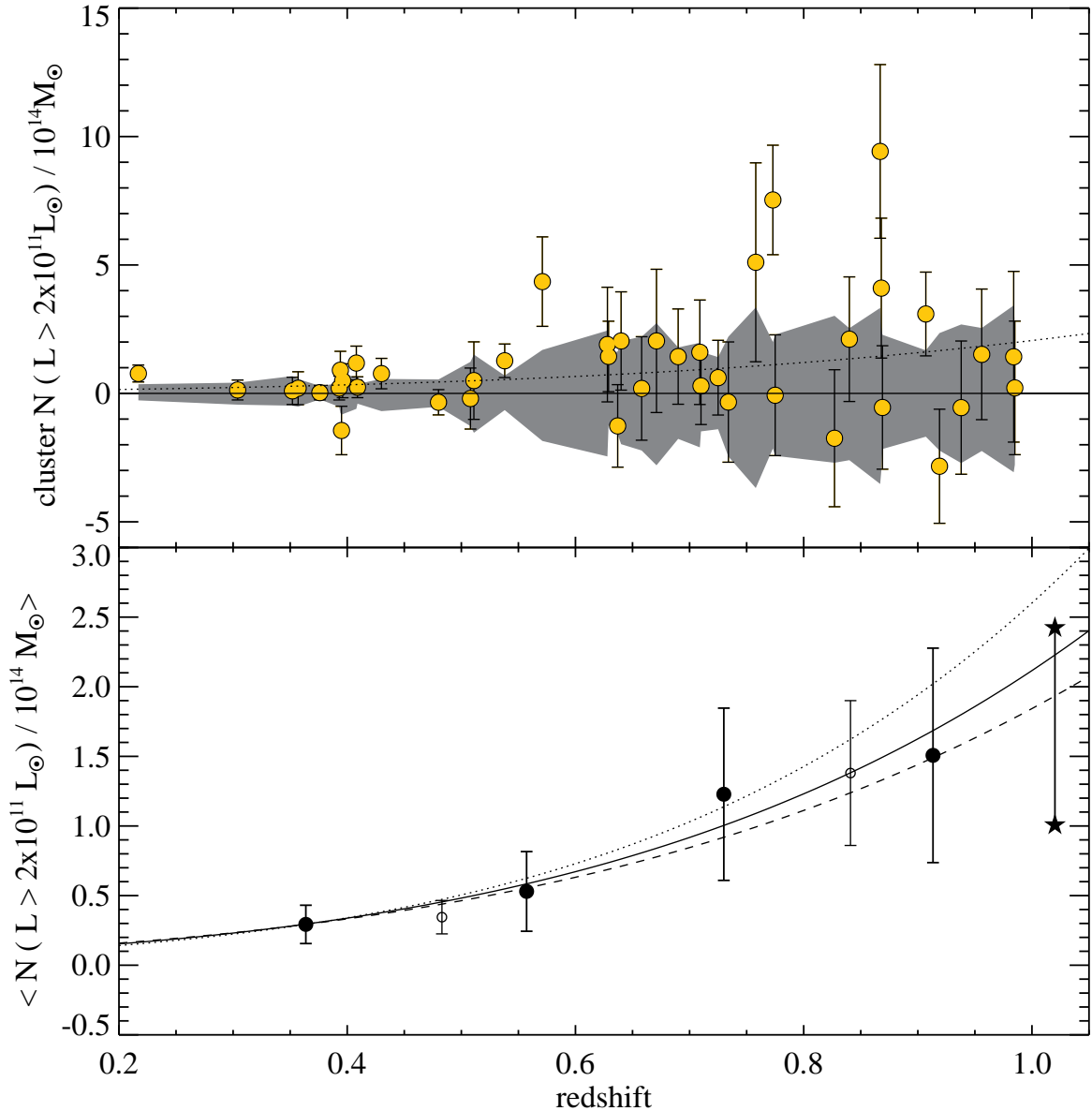}
\caption{The mean number of objects within $r_{200}$ in excess of the background counts with L$_\mathrm{IR} \geq$ 2$\times$10$^{11}$ L$_\odot$. The upper panel shows the measurements for each individual cluster.  Error bars represent the rms variation of the field counts and the blue band encompasses the 68\% region of the same analysis done on random locations in the COSMOS field. The  scatter in the mass-richness relation (see Appendix A) is not included here. The bottom panel shows the weighted average of the clusters in four redshift bins (solid points) and within two redshift bins (small open points), with the uncertainties corresponding to the standard error in the mean.   For the latter case the low and high redshift points are different at a level of 2.5$\sigma$. The top dotted line corresponds to the evolution of the fraction of star-forming galaxies in clusters  measured by \citet{haines09} at $z < $ 0.3 using  LoCuSS clusters and the data of \citet{saintonge08},  normalized to our measured value at $z =$ 0.4.      The middle black line shows the best fit power-law to the  four  RCS bins, of the form N($>$ S) $\sim$ (1+$z$)$^n$:   $n = $ 5.1$^{\pm 1.9}$, and this is also shown in the upper-panel (dotted line). The bottom dashed line shows the trend when the highest cluster in each bin is removed from the analysis. The two stars at $z\sim$ 1 denote the results of the  extinction-corrected optical  study of GCLASS clusters \citep{muzzin12}:  the lower  star corresponds to the confirmed number of star forming galaxies above 30 M$_\odot$yr$^{-1}$ (using the [OII] line measurements) and the upper star shows the value corrected for spectroscopic completeness (see discussion in text for more details).  The dotted line in the upper panel is identical to the solid line (best-fit function) in the lower panel. \label{counts} }
 
 %Solid points correspond to all clusters, small and large open points to clusters with Bgc $<$ 900 and Bgc $>$ 900 respectively.   As in \ref{fig 1} the dotted line denotes the trend resulting from a systematic over-estimate of Bgc (30\% in the highest redshift bin).  The dashed lines illustrate the difference in counts which would result from erroneously separating clusters of equal richness into two richness bins (see text).  Finally, the solid line corresponds  to an increase in IR luminous galaxies following the relation N($>$ LIRG) $\sim$ (1+$z$)$^n$ with the best fit value of $n = $ 4.9 $\pm$ 1.3.   Similar fits to the lower and higher richness clusters give 5.4$\pm$2.1 and 4.9$\pm$1.6 respectively; we therefore see no significant difference between these two groups.
 \vskip 0.5cm
 \end{figure*}

In Figure \ref{counts} we show the total number of galaxies with  L$_\mathrm{IR} \geq$ 2$\times$ 10$^{11} $L$_\odot$ in excess of the average background counts, within $r_{200}$ of each cluster center. We choose $r_{200}$ as the largest radius for which all clusters have uniform MIPS coverage. To remove additional scatter due to the random variation in the average mass from bin to bin we further normalize each integrated count by the M$_\mathrm{200}$ mass of the cluster (inferred from the richness as outlined in \S 2.1).   

The top panel shows the measurements of each individual galaxy cluster  and the  scatter from cluster to cluster is substantial; given the size of the error bars this is dominated by the variance in the background, though real differences between the clusters must also be important. The increase in scatter with redshift is expected as a fixed luminosity limit corresponds to a deeper flux limit at higher redshift and the angular size corresponding to a given $r_{200}$ decreases, therefore  the shot noise in the background  increases.  The blue region denotes the 68\% range in  measurements conducted on random areas of the COSMOS image.  In spite of the limitations inherent in statistical background subtraction, a clear increase in the average number of IR luminous galaxies per unit cluster mass is seen with redshift (bottom panel). To quantify the evolution we fit a simple power-law,  N ($>$ S) = N$_\mathrm{o}$ (1+$z$)$^n$,  and find $n = $ 5.1$\pm$1.9, where the uncertainties correspond to the 1$\sigma$ significance region in a numerical chi-squared fit.    We also show the resulting fit if the cluster with the highest value within each bin is removed (i.e. 10\% of the sample), as the lower dashed line.   In this case the evolution goes as $n=$4.8$\pm$2.0, indicating that the trend is not due to a single outlier in each bin.   Finally,  we note that the chosen power-law function is arbitrary, and though it allows a quantification  of the average evolution smoothed over redshift that can be compared to other work, it should not be interpreted as the evolution of a single cluster.  Indeed, as mentioned below, the true evolution may be much more stochastic as, for example, groups are accreted onto clusters.   To further illustrate this we simply show the data in two redshift bins (open points) where the difference between high and low redshift (divided at $z =$ 0.7) is 2.5$\sigma$. 

The absolute counts of IR galaxies and the rate of evolution, is dependent on the adopted SED.  As outlined above, we have used the now-standard libraries of Chary \& Elbaz and Dale \& Helou, but these do not include AGN contributions to the MIR emission.  If there is significant AGN contamination then we are in fact measuring the number counts to a different intrinsic flux level (lower L$_\mathrm{IR}$ for a given observed 24$\mu$m flux), and that flux level may not be constant with redshift, in particular if the rate of AGN contamination also evolves.  This is explored in more detail in the Appendix, where we conclude that we do not see evidence for variable contributions of AGN in our sample - though the statistics are poor.  Still, several studies have indicated that AGN are not a major contaminant over the  redshift and flux range explored here. For example, \citet{brand06} and more recently \citet{kirkpatrick13} conclude that at the 24$\mu$m flux levels of interest here  (200-600$\mu$Jy), significant AGN are present in only 10\% of galaxies.  Brand et al.~also estimate the amount of contamination AGN contribute to the total MIR flux to be $\sim$10\% for similar depths.  Finally, we note that because our technique relies on background subtraction of the statistical field, systematic differences between the two (as seen here) would require corresponding differences in the IR  SEDs of field and cluster galaxies - and would therefore remain an interesting result - though requiring a different interpretation.   Understanding the importance of AGN is extremely important and will be addressed in later work using more appropriate data than presented here.
 
\begin{figure}
\hskip -1cm \includegraphics[scale=0.5]{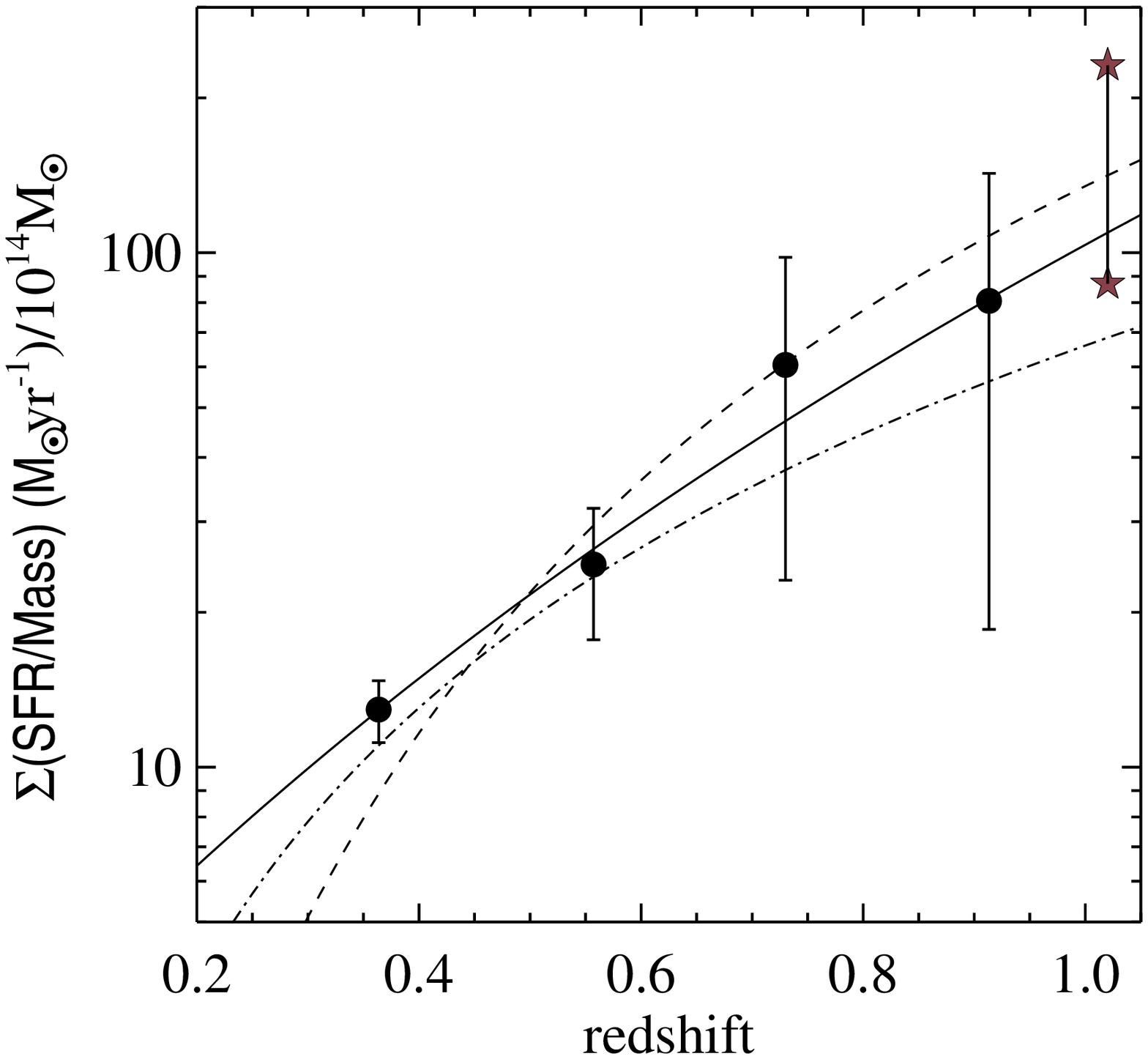}
%\vskip -4cm
\caption{The total star formation rate per unit cluster mass, $\Sigma$(SFR)/M, shown averaged over all clusters in four redshift bins. The solid line shows a fit to these data of the form   $\Sigma$SFR/M $\propto (1+z)^n$ with the best fit value $n$ = 5.4$\pm$1.5. The dot-dashed line shows the results of the similar study by \citet{popesso12} using 9 clusters over the same approximate redshift range, and the dashed line is the same scaled to our IR depth and average cluster mass. The two  stars correspond to \citet{muzzin12}, as in Figure \ref{counts}. \label{ssfr}}
\end{figure}

 For the remainder of the paper, we assume that the MIR emission is produced entirely by star formation.  In doing so, 
 these measurements can further provide an estimate of the evolution of the more physically meaningful quantity, the  total SFR per unit cluster mass, $\Sigma$SFR/M$_\mathrm{cluster}$ and this is shown in Figure \ref{ssfr}).  To estimate this we use the number counts of Figure \ref{counts} to set the normalization of the IR luminosity function \citep{lefloch05} at each redshift. We then estimate the total L$_\mathrm{IR}$ from galaxies above  2$\times$10$^{11}$L$_\odot$ by integrating the luminosity function.   We then convert this to a $\Sigma$SFR following \citet{bell03}.   We again fit a simple power-law and find roughly the same exponent as before: $n = $ 5.4 $\pm1.5$. 
 
%\subsection{The Specific SFR of clusters with redshift}

\subsection{Comparison with the Field}

We see a clear increase in the average number of  IR-luminous galaxies in clusters with redshift,  but this evolution must be compared with the behavior of field galaxies at similar redshifts.   Since clusters are continually accreting from the field any change in their galaxy populations with time may simply reflect the corresponding evolution in field galaxies feeding the infall, rather than cluster specific evolution.  

%Given that  strong evolution in number density of  the IR-luminous field population is now well established  \citep{lefloch05} it would not be surprising to see this mirrored in clusters. 

\begin{figure}
%\vskip -3cm
\hskip -1cm \includegraphics[scale=0.5]{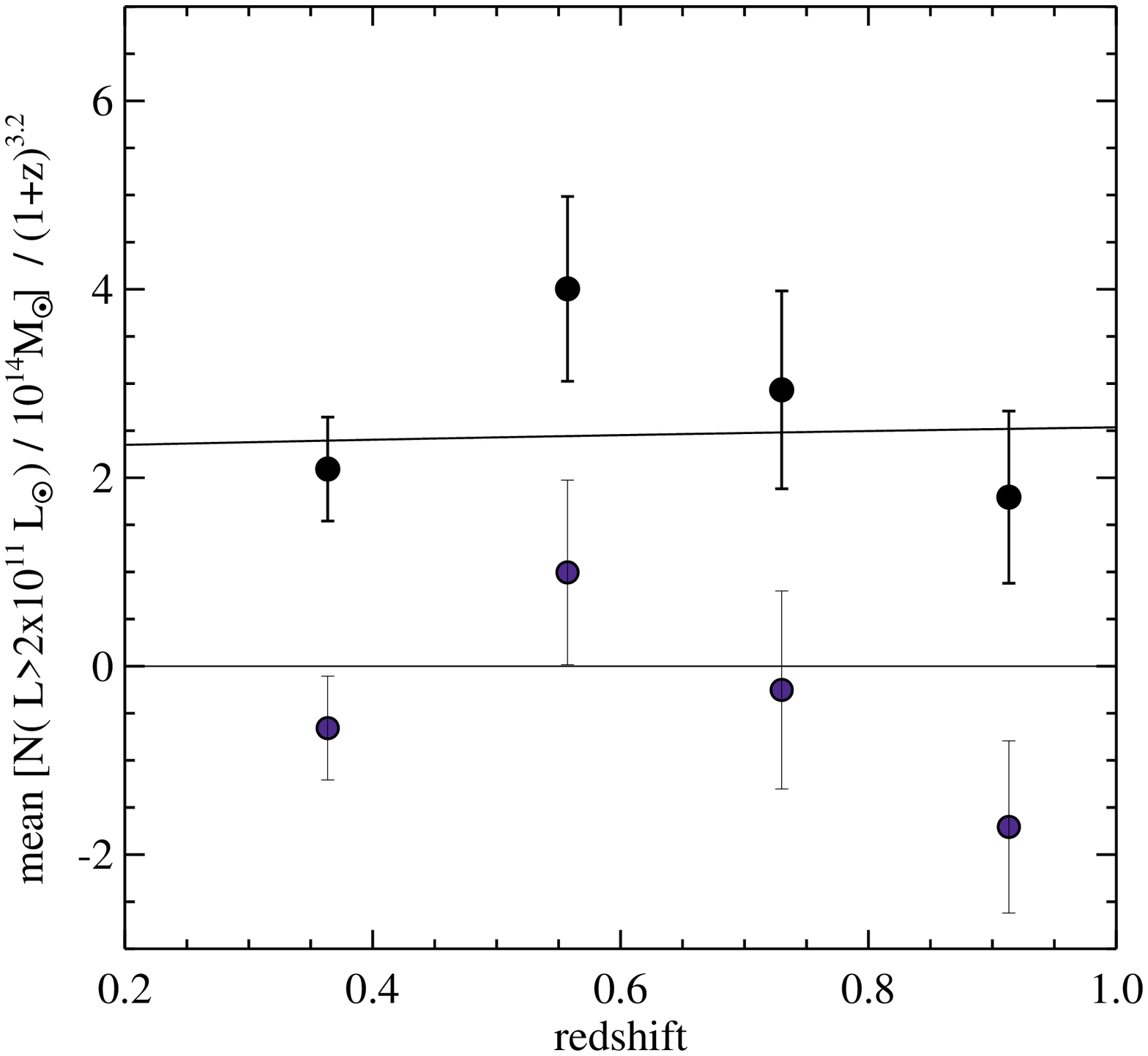}
\caption{Similar to Figure \ref{counts},  but with  the evolution of the IR luminosity function derived from field galaxies accounted for in the IR count depth. We measure the number of  IR luminous galaxies to the same depth relative to an evolving L$_\mathrm{IR}^\star$. The black solid line shows the best fit power law: N$\sim$(1+$z$)$^n$ to the data yields  $n = $ 0.5$\pm 1.4$. The blue points show the cluster values after the mean number of IR-luminous galaxies per halo mass in the field has beens subtracted; see \S 4.2 for details. \label{fieldevol}}
\end{figure}

To look for differences between field and cluster population evolution we use the same method of the lower redshift study of \citet{haines09}.  Instead of measuring the number of galaxies within clusters to a fixed luminosity at all redshifts, we vary the intrinsic depth to reflect the evolution in the IR luminosity  function of \citet{lefloch05}: L$^\star_\mathrm{IR} \sim$ $(1+z)^{3.2}$. We normalize to the previous depth of L$_\mathrm{IR} = $ 2$\times$ 10$^{11}$ L$_\odot$ at $z = $ 1 and show the corresponding flux depth with redshift in Figure \ref{depth}.  This has the added advantage that it allows us to reach the intrinsically deeper luminosity limits in the lower redshift images.  The results of this analysis are shown in Figure \ref{fieldevol} (upper black points).    An excess of IR bright galaxies is seen in all redshift bins,   reflecting the  over-density of galaxies at the location of the clusters, but there is now no evidence for an increase in the number  with redshift. A formal fit of N = N$_\mathrm{o}$ (1+$z$)$^n$ to the data yields  $n = $ 0.5$ \pm 1.4 $, indicating the cluster IR population  follows the  field evolution.  

For further comparison, we estimate the number of IR galaxies per unit 10$^{14}$M$_\odot$ in the field by integrating the luminosity function of \citet{lefloch05} at each redshift, and normalizing by the average matter density of the universe. Note that this is different than the {\it background} subtraction which removes galaxies along the line of sight, and the correction form the field IR luminosity function, which removes overall expected evolution of galaxies.  This simply scales the star formation per unit mass in the field to the mass of the cluster.   The difference between this field estimate and the cluster measurements is shown in blue and does not indicate an overall excess or deficit in the number of IR bright galaxies in clusters, once normalized for the halo mass. Note, however, that  while we have normalized the cluster counts to the mass within $r_{200}$  we may actually be sampling a larger mass along the line of sight, by up to 30\%, if the clusters are not highly concentrated \citep{lokasmamon01}. If so, we have then under-subtracted the field and the blue points should be lower, by roughly the same percentage, possibly indicating a small deficit of IR galaxies within clusters.
While several studies \citep[e.g.][]{kocevski11,marcillac07} claim to measure enhanced star formation in $z\sim$1 clusters, relative to the field, they have neglected the crucial step of accounting for the overall increase in mass (stellar or dark matter) in clusters.  Here we show that the overall star formation per unit halo mass, or per unit galaxy density, is either consistent with or lower than the field value; that is, {\it any apparent increase in star formation within this sample of clusters at 0.5 $<z<$ 1.0  is accounted for simply by the overall increase in galaxies at that location. }

% We note, however, that there are a number of uncertainties in this comparison.  Firstly, while we have normalized the cluster counts to the mass within $r_{200}$  we may actually be sampling a larger mass along the line of sight, by up to 30\%, if the clusters are not highly concentrated \citep{lokasmamon01}. If so, we have then under-subtracted the field and the blue points should be lower.   Secondly, our masses are calibrated on velocity dispersion measurements (Ellingson et al., in preparation) and \citep{white10} has shown that in simulated clusters the velocity dispersion could over estimate the total mass by up to 40\%.   While Ellingson et al. (in preparation) sees no evidence of this, These systematics do not alter the comparison between cluster samples, but could overestimate the number of IR galaxies, per unit mass, in clusters compared to the field. We therefore show these points as upper limits.

% Thus, the increase in LIRGs within clusters at earlier epochs can be explained entirely by infall from the evolving field galaxy population.   While not significant, it is interesting to note the slight excess in number counts at $z < $0.7. If true, this would imply a shallower evolution of L$_{IR}^\star$ from $z = $ 1 for clusters, compared to the field. 

\subsection{Correlation with Cluster Richness/Mass}

\begin{figure}
\hskip -1cm \includegraphics[scale=0.5]{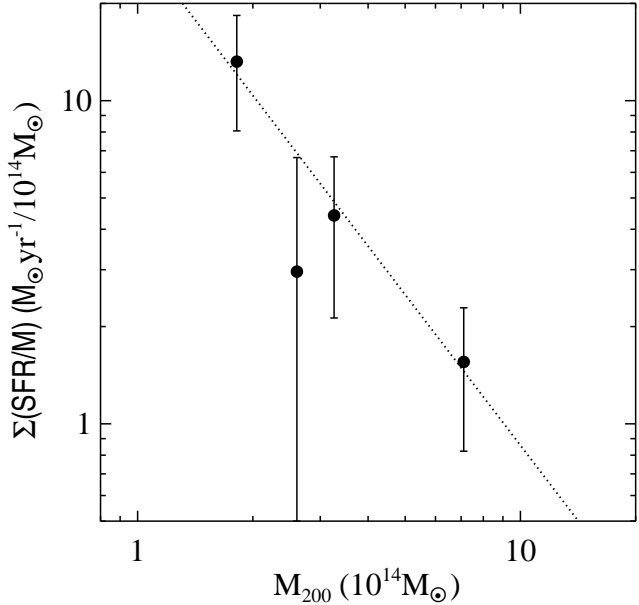}
\caption{The integrated star formation rate of clusters per unit cluster mass as a function of total cluster mass. As elsewhere, this includes all galaxies above L$_\mathrm{IR} > $ 2$\times$ 10$^{11}$ L$_\odot$ and within $r_{200}$.  The richness-mass conversion method is described in the text.   We have removed the effect of redshift evolution by normalizing all cluster counts to the  $z = 0.4$ level, following the relation measured in \S 4.1.   The dotted line corresponds to the  best fit power-law of $\Sigma$SFR/M$_\mathrm {cluster} \propto$ M$^{-1.5\pm0.4}$.
   \label{richness}}
\vskip 0.5cm
\end{figure}

%\begin{figure}
%\epsscale{1.2}
%\plotone{counts_r200_2d11_m200_chary_richness.ps}
%\caption{The mean number of objects within $r_{200}$ in excess of the background counts with $L_\mathrm{IR} \geq$ 2$\times$10$^{11}$ L$_\odot$,  normalized by the total cluster mass. The upper panel shows the measurements for each individual cluster; blue points correspond to clusters with Bgc $>$ 900 Mpc$^{-1.8}$ and red points to Bgc $\leq$ 900 Mpc$^{-1.8}$. The bottom panel shows the weighted average of the clusters in four redshift bins with  the blue and red points corresponding to the same richness division as in the top panel.   The solid lines shows the best fit power-law to the RCS data, of the N($>$ S) $\sim$ (1+$z$)$^n$:   $n = $  4.4$^{+0.7}_{-1.1}$; 6.5$^{+0.5}_{-1.4}$ for the low and high richness samples respectively. \label{counts_mass} }
%\end{figure}
In Figure \ref{richness} we show the integrated star formation rate per unit cluster mass, in four mass bins. As before, this includes all objects with L$_\mathrm{IR} >$ 2$\times$10$^{11}$ L$_\odot$ and within $r_{200}$, with the background subtracted.  We have further normalized all counts to the counts of $z= $ 0.4 to remove the redshift evolution, as we do not have enough clusters to bin in redshift and richness, and still maintain decent statistics. This last step is not strictly necessary provided all mass bins represent the same redshift distribution, but doing so should reduce the scatter due to small numbers.  We use dynamic bin sizes to ensure equal numbers of clusters in each bin;  as the mass distribution is not uniform over the range, this results in a larger bin width for the highest masses.   We fit a simple power-law and find  
 $\Sigma \mathrm{(SFR)/M}_\mathrm {cluster} \propto$ M$_\mathrm{cluster}^{-1.5\pm0.4}$.  
%  For comparison we draw  the \citet{bai09} relation of $\Sigma \mathrm{(SFR)/M}_\mathrm {cluster} \propto$ M$^{-0.9}$ scaled to roughly fit our results.  

As discussed in the Appendix, systematics in the mass normalization could bias the total SFR per unit cluster mass.    In Figure \ref{richness} however, systematic errors are not required to produce a trend; a false decrease in the integrated SFR per unit mass with cluster mass could be introduced through statistical uncertainties or through the intrinsic scatter ($\sim$30-40\%) in the mass-richness relation \citep{rozo11,rykoff12}.  Consider a sample of clusters of equal true mass but with a 30-40\% random error on their richness measurements. This would spread the sample non-uniformly over a larger  mass range (though not as large as explored here) and when binned in mass this results in a systematic over estimate of the mass in the high-mass bins, and a systematic underestimate of the mass in the low-mass bins. As discussed before this will result in an increase of $\Sigma SFR/M_\mathrm{cluster}$ at lower masses and a decrease at higher masses, as is seen. Still, once again, the measured slope is steeper than what can be accounted for by reasonable uncertainties alone.    For a flat radial distribution of SF galaxies, the expected slope relation is  $\Sigma$ SFR/M$_\mathrm{cluster} \propto$ M$^{-0.3}$.   Reproducing the observed relation requires an unreasonably steep radial distribution of galaxies (so that the number counts are dominated by the inner regions and no new counts are incorporated as the radius increases) and uncertainty on the richness measurements of more than a factor of two.  The decrease in the integrated SFR per unit cluster mass with cluster mass therefore appears robust.

%This  raises the concern that these clusters are less reliable objects, perhaps with a higher rate of contamination from groups aligned along the line of sight.  However, there is no reason a priori to think so:  though these are the lowest richness clusters in our sample they have masses $\gtrsim$ 3$\times$10$^{14}$M$_\odot$  and many have been spectroscopically confirmed (Table 1) with velocity dispersions in line with their Bgc measurements. In particular, the two clusters with the lowest IR counts in this group have spectroscopically verified redshifts and one has a velocity dispersion of 686 $\pm$ 225 km/s.

% I WOULD LIKE TO INCLUDE A COMPARISON WITH VEL DISP HERE IF POSSIBLE.

\subsection{The Radial Distribution of MIPS Galaxies}

\begin{figure*}

%\hspace{1.5cm}
%\vskip 1cm
\includegraphics[scale=0.8]{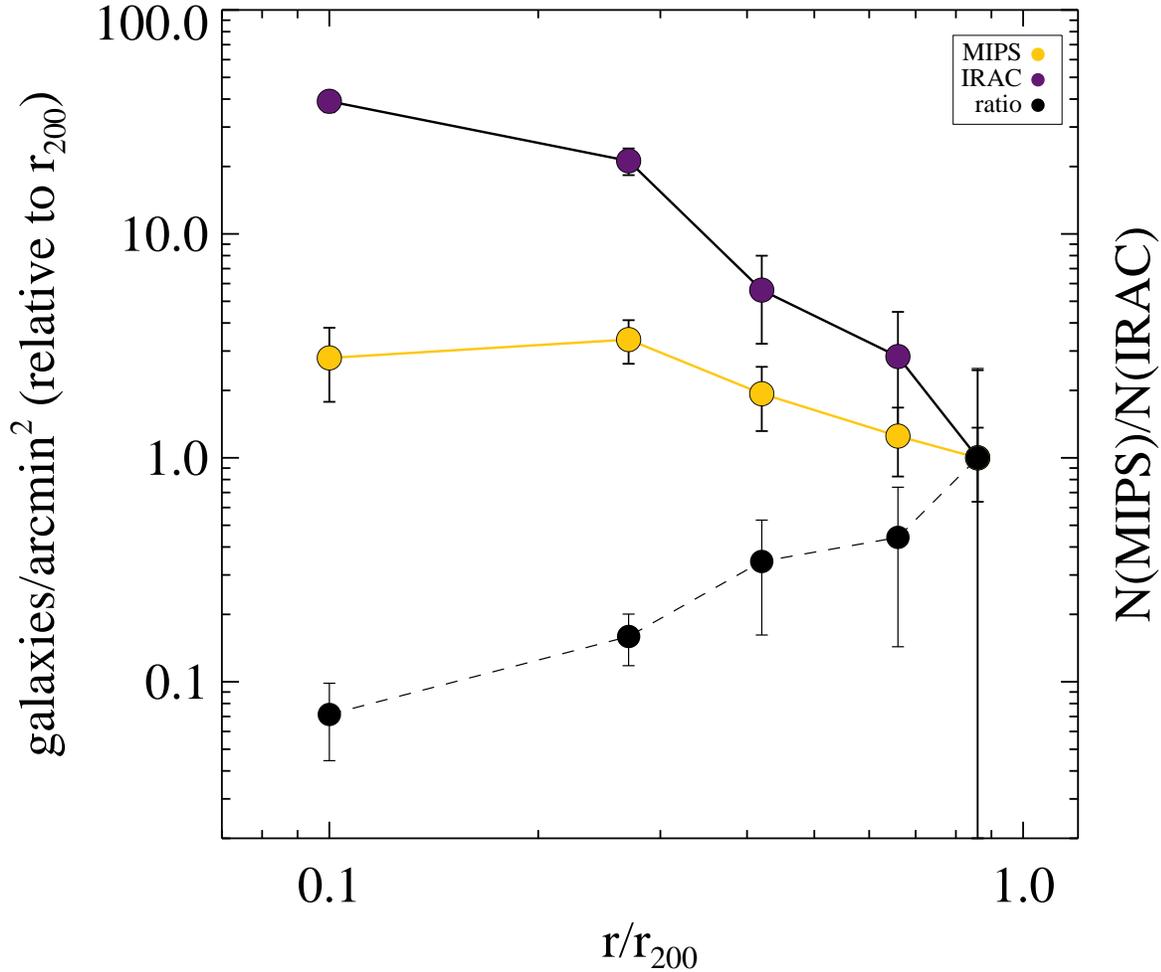}
%\vskip 1cm
\caption{The surface density of different galaxy populations in cluster-centric annuli.   The blue points/line correspond to IRAC detected galaxies with M$_\mathrm{K} > $ M$^\star$ and the yellow points/line correspond to the 24$\mu$m population with L$_\mathrm{IR} > $ 2$\times$10$^{11}$ L$_\odot/(1+z)^{3.2}$ (the dynamic limit applied in Figure \ref{fieldevol}.  Both populations have been background subtracted and normalized to the outermost radius. Each radial point represents a weighted mean over all clusters above $z = $ 0.5 (these are the only clusters with IRAC coverage).  The black points show the number of IR-luminous galaxies per unit area per unit IRAC galaxy, again scaled to the outermost bin.        \label{radialfig}}
\vskip 0.5cm
\end{figure*}

In Figure \ref{radialfig} we show the average number density of galaxies above the background in distinct cluster-centric annuli. We show the distribution for IR luminous galaxies (yellow) and perform the same analysis on the IRAC NIR galaxy population at the same locations in the cluster fields (blue).  Both are normalized to the outermost radial bin.  The NIR galaxies provide a trace of the underlying  galaxy surface density, and roughly correlate with stellar mass.  Note however, that the actual stellar mass density will be a steeper function of radius 
because the average stellar mass per galaxy increases toward the centre of the cluster \citep[e.g.,][]{muzzin12}.  We do not split the sample into redshift bins as the statistics become too poor:  thus there is the implicit assumption that the radial distribution is the same over 0.5$<z<$1.0, even though the total number of IR  galaxies evolve.  
Each population is analyzed separately and we do not require galaxies to be detected in both IRAC and MIPS. Figure \ref{radialfig} shows that the MIPS galaxies follow a relatively flat radial distribution, with  a hint of an increase  toward smaller radii,   reflective of the general increase in galaxy density in this region, as illustrated by the steep radial distribution of NIR-detected galaxies.  The black points show the ratio of the number density of these two populations -- the number of IR-luminous galaxies per unit rest-$K$-band selected galaxy (to the limits given in the caption).

%In the lower panel of Figure \ref{radialfig}   
%we show the average number of 24$\mu$m-detected galaxies per NIR galaxy.  This is similar to the fraction of optical galaxies  which are  star forming  (assuming no AGN contribution)  ($f_\mathrm{SF}$), but as we do not require the IR galaxies to have optical counterparts  a significant population of optically faint IR galaxies would increase this ratio. We see that even though the absolute number of IR luminous galaxies increases towards the centres of clusters, there is a sharp decrease when one accounts for the overall increase in galaxy density.  The ratio of IR to optical/NIR galaxies decreases by approximately an order of magnitude from 0.7$r_{200}$ to 0.1$r_{200}$.  

\section{Discussion}

\subsection{The evolution of the IR-luminous galaxy population in clusters}

Here we present the largest single study which attempts consistency over the redshift range 0.3 $<z<$ 1.0 in cluster selection, member identification and SFR measurement method.  
We find the absolute number of IR galaxies in clusters, normalized by the parent halo mass,  or the corresponding  approximate SFR/M$_\mathrm{cluster}$     rises as $(1+z)^{\sim 5}$.
The primary result of this work is that although there is seemingly rapid evolution, such behaviour is entirely consistent with the evolution of the IR bright field galaxy population over the same redshift range.  Thus the evolution of the IR galaxy population in clusters, and the inferred SFRs within clusters,  can be attributed solely to the change in the in-falling field population. 
%We see no evidence for any global enhancement of the star formation rates of galaxies during the final stages of the accretion process; that is, the SFR per unit optical galaxy (averaged over the whole population) is not in excess of the field value inside of $r_{200}$.   This does not rule out models where enhanced  star formation or nucleic activity occur outside of $r_{200}$, during earlier infall stages.  It is also blind to any process which averages out over the entire population.  For example, one might imagine a period of rapid but perhaps mild enhancement of star formation,  followed by strong quenching of this activity.  Averaged over many galaxies within $r_{200}$, the enhancement may not be visible. 

Given the variety of approaches and samples outlined above, it is difficult to directly compare our results to those of other authors.   The observations of single or a few clusters \citep[e.g.][]{kodama04,geach06,saintonge08} certainly agree very generally with our work, given that they reach different SFR limits, measure the total SFR to different radii, and select different cluster masses; but a meaningful comparison must be limited to a few surveys  which target larger numbers of clusters and use methods which are easily corrected to our own. The Haines et al. survey of 30 LoCUSS clusters below $z = 0.4$ (Figure \ref{counts}) is a large sample of systematically selected X-ray clusters, of a similar mass range as explored here. They undertake a statistical estimate of the fraction of star forming galaxies in clusters, relative to the optical population,  which is  analogous though not identical to our measurement, but reach 2$\times$ deeper in IR luminosity. They find a very similar rate of evolution as found here,  of $(1+z)^{5.7}$, which implies a continuation of the higher redshift evolutionary trend to $z\sim$ 0. 
 
 Our agreement with the recent results of \citet{popesso12}, shown in Figure \ref{ssfr}, is also heartening.  They reach 2$\times$ deeper in SFR (also IR-estimated) and thus their SFRs must be corrected upwards (using the LIR luminosity function at each redshift).  The average cluster mass of the Popesso et al. sample is also $\sim$3$\times$ higher than the RCS-MIPS average mass.  As we showed in the previous section (and found by others e.g. \citep{bai07}) the  $\Sigma$SFR/M$_\mathrm{cluster}$ decreases with cluster mass, and therefore the Popesso et al. relation must be scaled downward (we use the relation found in \S 4.4).  These two effects roughly cancel and preserve the good agreement between these two studies.
 
 Finally, in Figure \ref{counts} we also show  the $\Sigma$SFR/M of the GCLASS $z\sim$1 clusters \citep{muzzin12}.  The GCLASS clusters are derived from the SpARCS survey \citep{muzzin09,wilson09} and are also selected through the Red-Sequence technique; they therefore make an ideal higher redshift comparison.  The star formation is optically estimated through [OII] lines and includes spectroscopically identified cluster galaxies with [OII]  (extinction corrected), but we cut it at the same SFR limit as the MIPS data:  30 M$_\odot$yr$^{-1}$.  Again the agreement with the RCS number counts excellent (see Figure \ref{ssfr}).   The Popesso  and Muzzin programs differ from ours in a fundamental way in that the star forming galaxies must be detected in the optical. In the case of Popesso an optical counterpart is required and  for the GCLASS clusters the SFR is also estimated through the strength of the [OII] line. 
  The  agreement in the overall normalization of these three distinct measurements supports the idea that {\it optically biased studies are not missing a significant fraction of the star formation above this limit out to $z\sim$ 1}.   
 Still, this conclusion assumes that above a SFR of 30 M$_\odot$yr${-1}$, optical and IR observations are measuring the same population of galaxies, and this may not be the case:  optical studies will be biased to dust-free systems and IR observations to dust-enshrouded objects, and each may be missing a signifiant fraction of the other.   

\subsection{IR-Luminous Galaxies and Cluster Mass Assembly}

But how important is this in-falling IR population to the assembly of the stellar mass in clusters at $z\sim$ 0, and can we constrain the timescale over which this activity is shut down?  To investigate this we take an average mass halo in our sample, M$_{200} \sim$ 4$\times$10$^{14}$ M$_\odot$, and note that it will evolve into a present-day halo of mass $\sim$10$^{15}$M$_\odot$  \citep{wechsler02}.   By integrating the SFR of the IR-bright population over the redshift range probed here (taken from Figure \ref{ssfr}) we estimate that this population would add $\sim$4$\times$10$^{11}$M$_\odot$ of stars to the cluster.   If the decline continues with the same form to $z = $ 0 the mass added increases to $\sim$8$\times$10$^{11}$M$_\odot$.  Observations of local clusters indicate that the  stellar mass is $\sim$1\% the total halo mass \citep{andreon10}, or 10$^{13}$M$_\odot$ and therefore this simple treatment indicates the in-falling  infrared bright population below $z \sim$ 1 is responsible for 10\% of the total stellar mass of such a cluster today. 
However, the results of \S 4.4 indicate that as a cluster grows in mass its integrated SFR per unit mass will decrease following Figure \ref{richness}.  If we account for this additional shutting down of star formation, due not to the overall change in the in-falling population but to the change  in the parent halo  (again using the halo growth models of \citet{wechsler02}) the assembled stellar mass is reduced by roughly a factor of two, or to 5\%. 

The descedent population is harder to constrain. The only galaxy population which shows strong evidence for evolution from $z\sim$ 1 is the faint-end of the red-sequence 
\citep{gilbank08,delucia08,vulcani10}, however the estimated mass added by the IR-bright population is comparable to the {\it total} mass of the dwarf population in clusters at $z\sim0$ \citep{bildfell12} and would therefore require stronger observed evolution in this population than is currently seen, even with the above caveats.   However, we note that the uncertainties in the mass increase in the bright-end of the red-sequence over the redshift range probed here are large enough that they can incorporate the mass added by the IR-bright population \citep{bildfell12}.

Because of our statistical technique we have no information about the average current mass in stars in these systems; however the average mass of  field LIRGs at $z\sim$ 0.5-1 is $\sim$0.5$\times$10$^{11}$ M$_\odot$ \citep{genzel10,elbaz11}, or sub-M$^\star$ \citep{vulcani12}.   
To restrict growth to the faint-end of the red-sequence (or $\lesssim$10$^{11}$ M$_\odot$) the star formation seen here (taken to be at least 30 M$_\odot$yr$^{-1}$)  must be quenched within $<$ 2 Gyr.  This could be accomplished through simple gas depletion \citep{balogh00,genzel10,mccarthy08}  provided   the replenishment of the gas reservoir is shut down once the galaxy enters the cluster environment.    Still, longer quenching timescales are easily accommodated  since as noted above this population need not be restricted to the faint end of the red-sequence. 
Understanding the fate and more detailed importance of the IR-luminous population in clusters will be explored in future work, with the aid of spectroscopy, for now we simply conclude that our results are in line with other evolutionary studies of cluster galaxy populations.

% This scenario also offers an explanation for the mass dependency of the $\Sigma$SFR/M (discussed below) as higher mass clusters are capable of stripping larger fractions of the hot halo than less massive systems, thus shutting down the SF at a higher rate. 

%The consistent excess in all four redshift bins indicates that the IR galaxies are true cluster members, and not field interlopers. 

%See Balogh et al. 2000 and Gill et al. 2005 for discussion of infalling galaxies: backsplash

%See also Cortese 2003 for discussion of quenching 

%Treu et al. 2003 discuss ram-pressure stripping - it works within 0.5 $r_{200}$.   Substantial fraction of $> r_{200}$ galaxies are backsplashed but show no signs of decreased SF activity.   - however ram works better for lower luminosity galaxies 

\subsection{The dependence of star formation on cluster mass} 

%DISCUSS CLUSTER RICHNESS DEPENDENCE HERE - WRT POGGIANTI, BAI, FINN, WETZEL

The decrease in $\Sigma$SFR/M$_\mathrm{cluster}$ with halo mass seen in Figure \ref{richness} is not surprising:  this behavior has been seen by others \citep{finn04,finn05,bai07} (but {\it c.f.} \citet{chung2012}) and indeed is expected from the properties of local (i.e. $z = $0)  galaxy clusters.   The stellar mass fraction of clusters decreases with increasing cluster mass \citep{andreon10} as does the metal  enrichment of the ICM , while total gas mass increases \citep{zhang11}.  This implies cluster star formation efficiency that is dependent on the halo mass, such that high mass halos turn a smaller fraction of their baryonic mass into stars.  This could be the result of a constant dependence of the  quenching efficiency of clusters on mass over all time, so that while  accreting more mass from the field, high mass clusters preferentially return the gas to the IGM rather than process it into stars. Additionally, this difference could reflect a change in environmental quenching mechanisms with time.  At a given epoch clusters with different halo masses have different formation histories, assembly times and galaxy population  ages.   N-body simulations, local stellar population studies, and high-redshift observations of clusters indicate that high mass clusters formed a larger fraction of their stars at earlier times \citep{thomas05} than lower mass systems, thereby imprinting temporal variations on global star formation efficiencies on the cluster galaxy populations. 

 Over the mass range probed here ($\sim$10$^{14-15}$M$_\odot$) local clusters show a M$_\star$/M$_\mathrm{halo}$  dispersion of approximately an order of magnitude. The simplest interpretation of this is that  their star formation efficiency follows the same ratio (with the  higher mass clusters an order of magnitude less efficient at forming stars than the lower mass systems) , at least averaged over all time. Our measured relation of the $\Sigma$SFR/M$_\mathrm{cluster}$ is consistent with this at $\sim$1$\sigma$, and therefore does not require a change in the quenching efficiency of star formation with redshift; that is, the difference between high and low mass clusters can be accounted for by the relation we see below $z\sim$ 1, and does not require differences to be set in place at early times.  According to these measurements, high and low mass clusters have assembled roughly the same amount of stellar mass since $z\sim$ 1, though this represents a smaller fraction of the total stellar mass of high mass clusters.  Thus, we confirm the fraction of a cluster's stellar mass that was formed at high redshifts increases with cluster mass. This is also in line with the richness dependent downsizing effect observed by \citet{gilbank08} (but \citep[c.f.][]{delucia08,bildfell12}:  lower mass clusters show a larger deficit of faint red sequence galaxies than high mass clusters to $z\sim$ 1.  
 
 %We note however, as a matter of interest, that the slope in Figure \ref{richness}, although consistent with $\Sigma$SFR/M$_\mathrm{cluster}\propto$ M$^{-1}$, is slightly steeper, and thus formal integration implies more stellar mass has been produced in 10$^{14}$M$_\odot$ clusters since $z\sim$ 1 than in 10$^{15}$M$_\odot$ systems. Taken at face value this implies that high mass clusters are in fact too efficient at quenching star formation  below $z\sim$ 1 and to account for the M$_\star$/M$_\mathrm{halo}$ relation of the local universe high mass clusters would have been {\it more} efficient at forming stars in the past.   Indeed, somewhere beyond $z\sim$ 1 the star-formation-rate density relation reverses \citep{elbaz07}, such that  while star-formation avoids dense regions in the local universe,  the ancestors of dense cluster cores were the preferred site for star formation in the deep past, and the efficiency of star formation in these regions might have been higher than today.  Although not required by our results, the uncertainties can accommodate this interesting possibility.  

The above interpretation is in line with the properties of low redshift halos, but it is not the only possible explanation for Figure \ref{richness}.  Disentangling the effects of environment and galaxy mass is notoriously difficult and could be important here.  The specific star formation rate of individual galaxies is known to decrease with increasing mass even to $z\sim$ 1 \citep{elbaz07} and thus if the mass function of in-falling  galaxies varies with parent halo mass, such that high mass clusters preferentially accrete high mass galaxies, this would be reflected in the overall specific star formation rate of the clusters. This scenario is not supported by the recent measurements of the mass function of galaxies surrounding clusters to  $z\sim$ 1 of \citet{vulcani12}, and would nevertheless be too weak an effect to produce what is seen here.
Still,  N-body simulations do highlight a significant difference in the accretion history of halos with mass.  \citet{mcgee09} show that high mass clusters form preferentially from group accretion, or the accretion of halos with M$>$10$^{13}$M$_\odot$, whereas lower mass systems primarily accrete isolated field galaxies.    In this scenario star formation is first suppressed by the local environment \citep{balogh11} before accretion onto a massive halo and the apparent dependence on the global environment is simply a reflection of the mass function of accreted halos and the larger number of groups falling into the highest mass clusters.

% It is also necessary to remind the reader that these conclusions assume a constant fraction of AGN with halo mass.  If this is an invalid assumption then Figure \ref{richness} cannot be simply interpreted as tracing the star formation efficiency of clusters with total mass. Nevertheless, it would still imply an interesting physical difference between the activity quenching and triggering mechanisms of halos of different masses.  In this case low mass clusters would contain a higher fraction of AGN emission in their integrated infrared light.  
 
 Finally, we recall that  the RCS clusters are optically selected with the mass estimated through optical richness measurements.  This means that our sample is fundamentally a stellar-mass limited sample, rather than a halo-mass limited sample.  This could introduce a systematic bias in stellar-to-total mass ratio with total mass. At the high mass end we will be complete for all possible ratios, but at low mass we could be biased to clusters with high stellar mass content, and therefore those systems which have experienced more efficient star formation for the same dark matter halo mass.   This would take a flat $\Sigma$SFR/M$_\mathrm{cluster}$ relation and tilt it negatively by biasing the low mass clusters upwards.  Such an effect should be recoverable provided one has an independent estimate of the total mass, which we currently do not.
  
 % While we see no evidence for such an effect in Figure 2 (where we have tied the optical richness measurements to velocity dispersions) the numbers of clusters are likely too small to be conclusive. 

\subsection{The radial dependence of the star formation beyond $z>$ 0.5}

Figure \ref{radialfig} confirms the  decrease in star-formation rate per unity galaxy toward high density regions - the so-called star-formation-rate density relation which has been well-established by many authors.  While field studies show that the relation begins to reverse beyond $z\sim$ 1 \citep{elbaz07,cooper08},  evidence of it reaching cluster core densities below $z\sim$ 1.6 \citep{kodama04,tran10,muzzin12} is not yet conclusive.   
As with the evolutionary effect discussed above, it is exceedingly difficult to directly compare the results of different studies. The absolute value of the fraction of star forming galaxies depends on the depth and method of the star formation rate estimates and the method of assessing the underlying cluster population.  Nevertheless, most studies see a smooth decline in the fraction of star forming galaxies of about  an order of magnitude from the outer regions of the cluster ($\sim r_{200}$) to the core \citep{kodama04,muzzin12,patel11}. It is now clear that some, but not all, of this trend is due to the underlying mass-bias \citep{muzzin12};  the average mass of galaxies increases towards the cluster core and the lower sSFR of higher mass galaxies drives down the average star formation rate.   

%In Figure \ref{radialfig} we compare our radial distribution to that of \citep{muzzin12}.  
%The red line and triangles denote the relation for [OII]-measured SFRs of spectroscopically identified cluster members in the $z\sim$1 sample of GCLASS clusters.  This measurement differs from our in a number of significant ways.  First, Muzzin et al. measure a true $f_\mathrm{SF}$ because they limit themselves to spectroscopically confirmed members, but this should not introduce any systematic difference between our samples.  Second, they reach  SFR $\sim$10 M$_\odot$/yr, compared to the   $\sim$ 30 M$_\odot$/yr  probed here.  It is straightforward to account for the differing SFR depths by applying a simple offset to the Muzzin et al. ratios.   We use the LIR luminosity function at $z \sim$ 1 to scale the counts of Muzzin et al. downwards.  This brings the two sets of measurements into agreement within the generous uncertainties of our statistical study in the outer two radial bins, but a small offset persist in the core. Finally, the Muzzin et al.  points shown here correspond to a limited range of stellar masses:  10.0 $< $log (M$_\star$/M$_\odot$) $<$ 10.7.  This matches the approximate depth of our comparison optical/NIR sample, but whereas we include all masses above M$^\star$+1, they work in a mass range and therefore do not include the highest mass galaxies.   This confirms the idea that the SFR-density relation is shallower for mass-controlled samples. 

The statistics of our study are already too poor to do much beyond  describe the overall decline (see caption of Figure \ref{radialfig}) in star formation in the cluster centers.  However, it would be interesting in future work to investigate the rate of decline for galaxy populations of different star formation rates or specific star formation rates as this would provide further information on the galaxy dependent properties of the  quenching mechanisms.

\section{Conclusions and Final Remarks}

The simple statistical exercise presented here leads to a number of important conclusions:

\begin{enumerate}
\item{We see an steep increase in the number of IR-luminous galaxies (L$_\mathrm{IR} > $ 2$\times$10$^{11}$L$_\odot$) per cluster mass (and by inference $\Sigma$SFR/M$_\mathrm{cluster}$)  of $(1+z)^{5.4\pm1.9}$ over the range 0.3 $<z<$ 1.0.  This evolution is in agreement with that estimated by other IR studies of galaxy clusters \citep{haines09,popesso12} and shows the same level of increase as seen in extinction corrected optical studies to similar SFR depths \citep{muzzin12}.  Assuming the optical population is contained within the IR population (this has not been demonstrated), this indicates that the optical studies are not missing a significant fraction of dust enshrouded activity.}
%and indicates that there is not a substantial amount of star formation in clusters that is missed by optical studies or biases.   which can be explained entirely by infall from the field}
\item{We show that the above rapid evolution can be accounted for entirely by the evolution in the in falling field population.  Moreover, we show that the amount of IR-traced star formation per unit halo mass in clusters and the field are consistent; once normalized for halo mass clusters do not show an excess or in star formation deficit relative to the field. }

\item{The IR luminous in-falling population seen here to $z\sim$1 can account for 5-10\% of the total stellar mass in massive clusters today. Until we have an estimate of the stellar mass function of these galaxies, it is not certain how much of this population contributes to the build-up of the faint-end of the red-sequence observed in clusters over this timeframe.}
\item{Averaged over all redshifts, we see a decrease in the $\Sigma$SFR/M$_\mathrm{cluster}$ with  increasing galaxy richness of $\Sigma$SFR/M$_\mathrm{cluster}\sim$M$_\mathrm{cluster}^{-1.5\pm0.4}$.  This means that the SFR of an individual halo decreases more sharply with time than the simple number count analysis implies.  The relation is consistent with a constant dependence of quenching efficiency (but does not require it) on halo mass over all time and can reproduce the dependence of the  stellar-to-total mass ratio on mass seen in local clusters.    }
\item{The radial distribution of IR galaxies in clusters is flat, with perhaps a slight increase towards the centers. This is seemingly in agreement with other studies which claim enhanced star formation in the centers of high redshift clusters. This, however, is not a correct interpretation; once the underlying galaxy density is taken into account (a step often neglected by other studies)  we  see a decrease in the number of IR-luminous galaxies toward the cluster core for $z >$ 0.5 clusters.    Thus, the SFR-density relation persists to  cluster core densities  to $z\sim$ 1. }
 
 \end{enumerate}
 
 As noted in the introduction  the broad statistical technique limits the conclusions to general statements concerning the average cluster and galaxy populations, and cannot
 provide information about the properties of the individual galaxies. Moreover, we cannot control for the important interdependencies of stellar mass and environment. Such
 work will be the focus of later papers by our team and will better elucidate the history, characteristics, and fate of galaxies recently accreted into the
 cluster environment.

\acknowledgments

T.W. acknowledges the support of the NSERC Discovery Grant 
and the FQRNT Nouveaux Chercheurs programs.   Support for K.C. is
provided, in part, by the Centre de Research en Astrophysics du Qu\'ebec,
a {\it regroupment strat\'egiques} of the FQRNT and the Lorne Trottier Chair in Astrophysics
and Cosmology.  J.G. is supported by the NSERC Banting Postdoctoral Fellowship program.
G.W. gratefully acknowledges support from NSF grant AST- 0909198. H.K.C.Y. acknowledges support from
and NSERC Discovery Grant and a Tire 1 Canada Research Chair. A.F. is supported by an NSERC Graduate Fellowship.
Finally, we thank the anonymous referee for providing helpful comments that improved the work.


%% in parentheses, after the keyword, but they will not be verified.

{\it Facilities:} \facility{Spitzer Space Telescope (MIPS; IRAC)},  \facility{Magellan (IMACS)}, \facility{CTIO}, \facility{CFHT}.



\appendix

 \section{Possible sources of systematic error in the number counts} 

In interpreting the apparent increase in IR galaxies in clusters with redshift, one must be wary of other systematic trends with redshift (inherent or introduced)  which might mimic such evolution.  
There are three primary sources of possible systematic error in this analysis: (i)  the background subtraction (ii)  the use of a single template SED in inferring the observed 24$\mu$m flux for a given IR luminosity limit and  (iii) the calculation of the $r_{200}$ radius.   We discuss each in turn, but conclude that none are sufficient to account for the trend we see. \\

\noindent {\bf Background Subtraction:} Our analysis method is identical in the cluster and COSMOS fields and thus a  redshift-dependent offset in the counts would require a systematic difference between the cluster and SCOSMOS fields that varies with source brightness.  
 A difference in completeness depths could mimic source density evolution; however, a number density excess at high redshift would require the SCOSMOS imaging to be shallower than the cluster fields and we have shown in Figure \ref{diff_counts} that this is not the case;  the adequate depth of the COSMOS data is further confirmed elsewhere  \citep{sand07}.  
 
 If, however, the SCOSMOS GO3 field is located in a region of the sky of true galaxy under density at all relevant luminosities, this could introduce a trend similar to that observed. Again, our analysis of the number counts using our own photometry indicates that the number density of 24$\mu$m sources in SCOSMOS is in good agreement with other fields, and this is confirmed by the very careful analysis of the SCOSMOS team \citep{lefloch09}.  
 
  A final possibility is an increased level of asteroid contamination in the RCS fields.   We have not removed asteroids in either the RCS or SCOSMOS images, thus if the RCS-MIPS fields contain a higher number of asteroids than SCOSMOS, with the number density of asteroids increasing with decreasing flux, this might mimic redshift evolution.  We have checked this by plotting the excess IR counts against the ecliptic latitude of each cluster and see no correlation.  Moreover, the SCOSMOS field lies at lower ecliptic latitude than any of our clusters and therefore should have {\it more} asteroids. 
  Therefore, it does not appear that our analysis method or inherent differences between the cluster and field catalogs are likely to create a false increase in IR galaxies in the higher redshift bins.  \\
 
\noindent {\bf Luminosity Calculations:} We adopt the Chary \& Elbaz  (2001) and Dale \& Helou (2002) libraries of IR luminous galaxy templates to predict an observed 24$\mu$m flux for  L$_\mathrm{IR} \geq$ 2$\times$ 10$^{11} $L$_\odot$ for each cluster  and are therefore applying the same SED to all redshifts.   The Chary \& Elbaz SED library has been demonstrated to accurately reproduce the total IR luminosity of field galaxies below $z < 1.4$, particularly for galaxies  with L $< $ 10$^{12}$ L$_{\odot}$  which dominate our number counts, even when extrapolated from a single 24$\mu$m measurement (Magnelli et al. 2009; Murphy et al. 2009; Chary 2010). Thus, field studies show no evidence of a strong evolution in the IR SED over this redshift range and we therefore do not expect this to be a substantial effect. 
Nevertheless, if  the average SED of IR galaxies within clusters changes with redshift we could be measuring the number counts to different intrinsic luminosities at different cosmic epochs. To reproduce the observed trend of increasing numbers of IR galaxies at high-redshift requires an enhancement of 24$\mu$m flux for a given L$_\mathrm{IR}$ in the higher redshift bins.  
An obvious means of accomplishing this is through increased AGN contamination as the presence of hot dust surrounding an AGN  would produce excess MIR emission thereby boosting galaxies above the luminosity limit.  

To investigate this possibility we have computed the IRAC MIR colours of the 24$\mu$m galaxies in each cluster field (to the corresponding luminosity limit);  the MIR has been used to roughly classify galaxies as possible dusty AGN \citep{lacy04} through their location on an IRAC color-color plot.   We measured the fraction of 24$\mu$m galaxies in the cluster fields which have colors consistent with AGN according to the Lacy et al. classification in three separate redshift bins above $z =$ 0.5 and find no change in this fraction with redshift.   However, given the small number of excess IR galaxies in each cluster field (2-20 galaxies per cluster), relative to the high source density of the background, color differences will be difficult to detect using this statistical method and the results are therefore not strongly conclusive. This will be investigated in later work using more extensive imaging and spectroscopy.   We further note that a change in the IR SED with redshift, due to increasing AGN or other physics, simply means the evolution in number counts cannot be interpreted as an increase in the total SFR; it is still a real and interesting evolutionary trend. \\

\noindent{\bf Adopted Radii:} Since we are counting galaxies within a given radius for each cluster and the density of IR galaxies themselves may not be constant with cluster-centric distance,  the accuracy of the analysis also relies on the adoption of a consistent radial cut, taken here to be  $r_{200}$, or the approximate virial radius.  We calculate $r_{200}$  from the observed galaxy richness N$_\mathrm{red}$, which is an indicator of mass reliable to $\sim$30-50\%  \citep[][Gilbank et al. in preparation]{rozo09,white10}. 
 We choose not to attempt to correct for incompleteness in the highest redshift bins by assuming a constant LF, since we know that the faint end of the red-sequence becomes increasingly depopulated towards higher redshift \citep[e.g.][]{gilbank08} in a way which approximately mimics incompleteness.

Systematic offsets in the measured N$_\mathrm{gal}$ with redshift could introduce a corresponding trend in the number counts.  
 For example, if the galaxy population in the cores of clusters evolves substantially between $0.2 < z < 1.0$ an optically estimated N$_\mathrm{red}$ is unlikely to have a uniform conversion to total cluster mass. In this case we would expect the same N$_\mathrm{red}$ to correspond to a more massive system at high redshift (because the galaxies are bluer), compared to low redshift. This would lead to an underestimate of $r_{200}$ and therefore a decrease in the corresponding number counts, but also an underestimate of the mass normalization, thereby boosting the measurement.  However, initial spectroscopic results (Gilbank et al. in preparation) do not show a significant effect.
 
%We are cognizant of a number of possible systematics, but none can reasonably account for such rapid redshift evolution.  Firstly, N$_\mathrm{red}$ is estimated to a uniform optical depth of M$^\star$+2 at all redshifts, however the RCS imaging data is incomplete to this depth at $z\sim$ 1.   This results in a underestimate of N$_\mathrm{red}$ at the highest redshifts of $\sim$\%. This therefore implies we are counting infrared galaxies to too small a radius, or {\it underestimating} the absolute IR counts.  The normalization mass however, will also be underestimated and will lead to a boosting of the counts per unit mass. Similarly,
  
  The counts and mass normalization depend on different powers of the richness measure, N$_\mathrm{red}$.  As outlined in \S 3, we determine $r_{200}$ by first transforming N$_\mathrm{red}$ to a velocity dispersion ($\sigma$), following $N_\mathrm{red} \propto \sigma^{1.9}$;  the radius is then linearly proportional to $\sigma$.  Assuming a flat radial distribution of IR galaxies (\S 4.5) the absolute number counts will increase with the area, or as $r_{200}^2$.  The mass is a slightly stronger function of radius, rising as $r_{200}^3$.   These competing dependencies result in a bias upwards in the number counts per unit mass, but the effect cannot reproduce the evolution for any reasonable error in richness.  For example, a systematic underestimate of the optical richness in the $z = $ 0.9 bin of $\sim$30\% would require a correction  of the counts  downward by $\sim$10\%,  preserving a strong evolutionary trend. 	 Thus a large systematic in richness measurements with redshifts can steepen the evolution, but cannot completely account for it.

\begin{figure}
\includegraphics[scale=0.8]{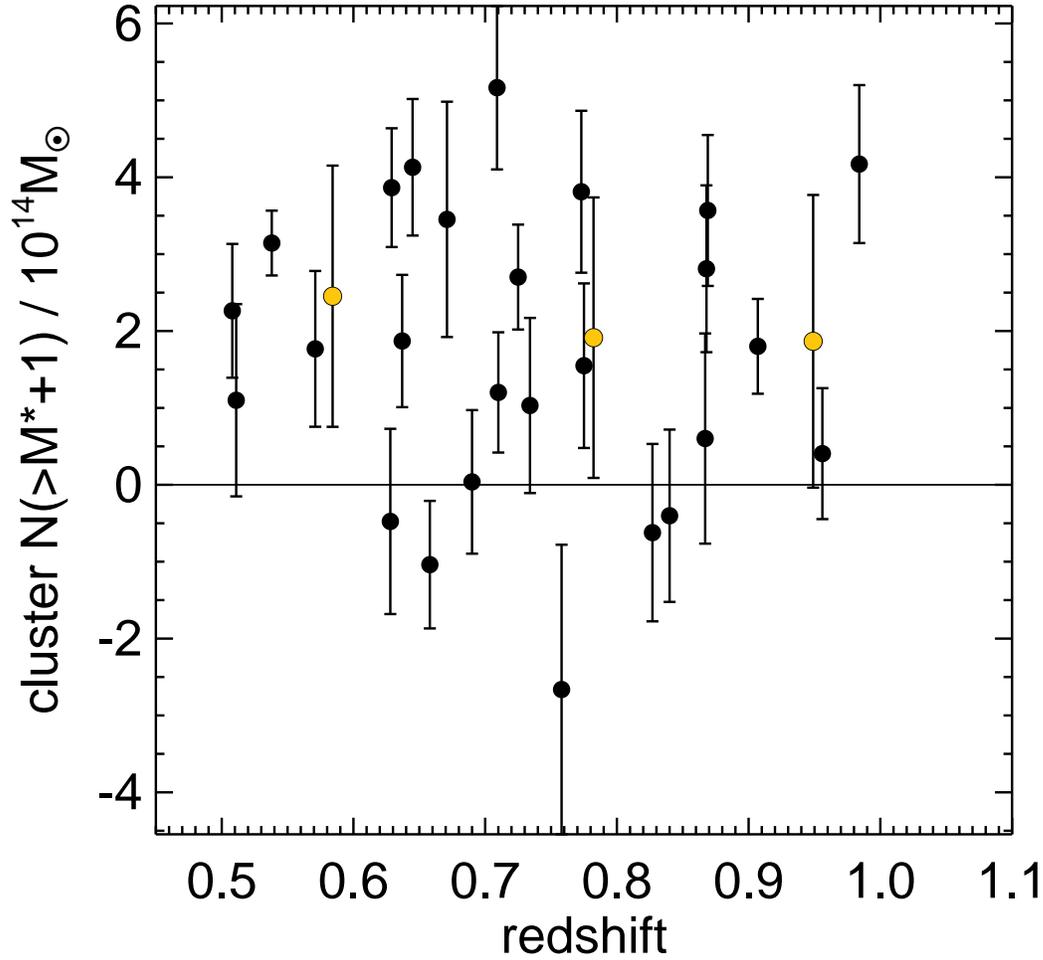}
\caption{The number of IRAC detected galaxies (3.6$\mu$m or 4.5$\mu$4m) within 0.5 $r_{200}$ and brighter than M$^\star$+1, after background subtraction.  The black points correspond to individual clusters and the yellow points show the weighted average within three redshift bins.  \label{irac}}
\end{figure}

 We have further checked this effect using the IRAC imaging for the $z>$ 0.5 sample.   We carry out a similar number count analysis as in the rest of the paper where  we measure the number of galaxies within a given radius and above a uniform luminosity limit, and perform background (line-of-sight) subtraction.  In this case we have chosen 0.5 $\times$ $r_{200}$, to maximize S/N,  and M$_\mathrm{K}$ $\gtrsim$ M$^\star_\mathrm{K}$ + 1, to remain well above the IRAC limiting depths at all redshifts. This is not a perfect reproduction of the MIPS analysis method since the optical/NIR galaxies are unlikely to have the same radial distribution as the IR galaxies (\S 4.4) at 0.5 $r_{200}$, and the luminosity conversion method is different, but it should provide a handle on the importance of large systematic errors in the counts due to systematics in the radii.  Within small radii, IRAC traces the stellar mass of the early-type galaxies dominating the core and should show very little evolution with redshift (in particular since these systems have been selected by the existence of an old optical population).    Thus, any large evolutionary trends in the IRAC counts may be attributed to systematic effects. 
 In fact, figure \ref{irac} shows no trend of increasing  optical galaxy counts with redshift, in stark contrast to the MIPS analysis.

\section{Raw Counts}

Here we show the background-subtracted 24$\mu$m counts in constant radii of 1Mpc and without normalizing for the inferred cluster mass. This plot makes no assumptions or corrections for the differences in counts which are due to differences in the cluster mass and does not infer an $r_{200}$ radius.  It should therefore be free from uncertainties introduced through errors in the assumed optical-richness-mass relation, or its intrinsic scatter.   On the other hand, working at a fixed radius and not accounting for mass should result in a larger scatter - assuming the corrections are properly applied.   This is because 1Mpc corresponds to a larger fraction of $r_{200}$ for poor clusters, compared to richer clusters, and because higher richness clusters may have larger numbers of all galaxy populations - including the IR-luminous systems.  The increased number of IR-luminous galaxies in higher mass clusters appears robust in that it is present in both the uncorrected raw data, and (at a higher level) for the corrected data.   For the mass-uncorrected data shown here the evolution goes as $(1+z)^n$ where $n=$4.4$\pm$1.9, compared to $n=$5.1$\pm$1.9 for the corrected version.

\begin{figure}
\includegraphics[scale=0.5]{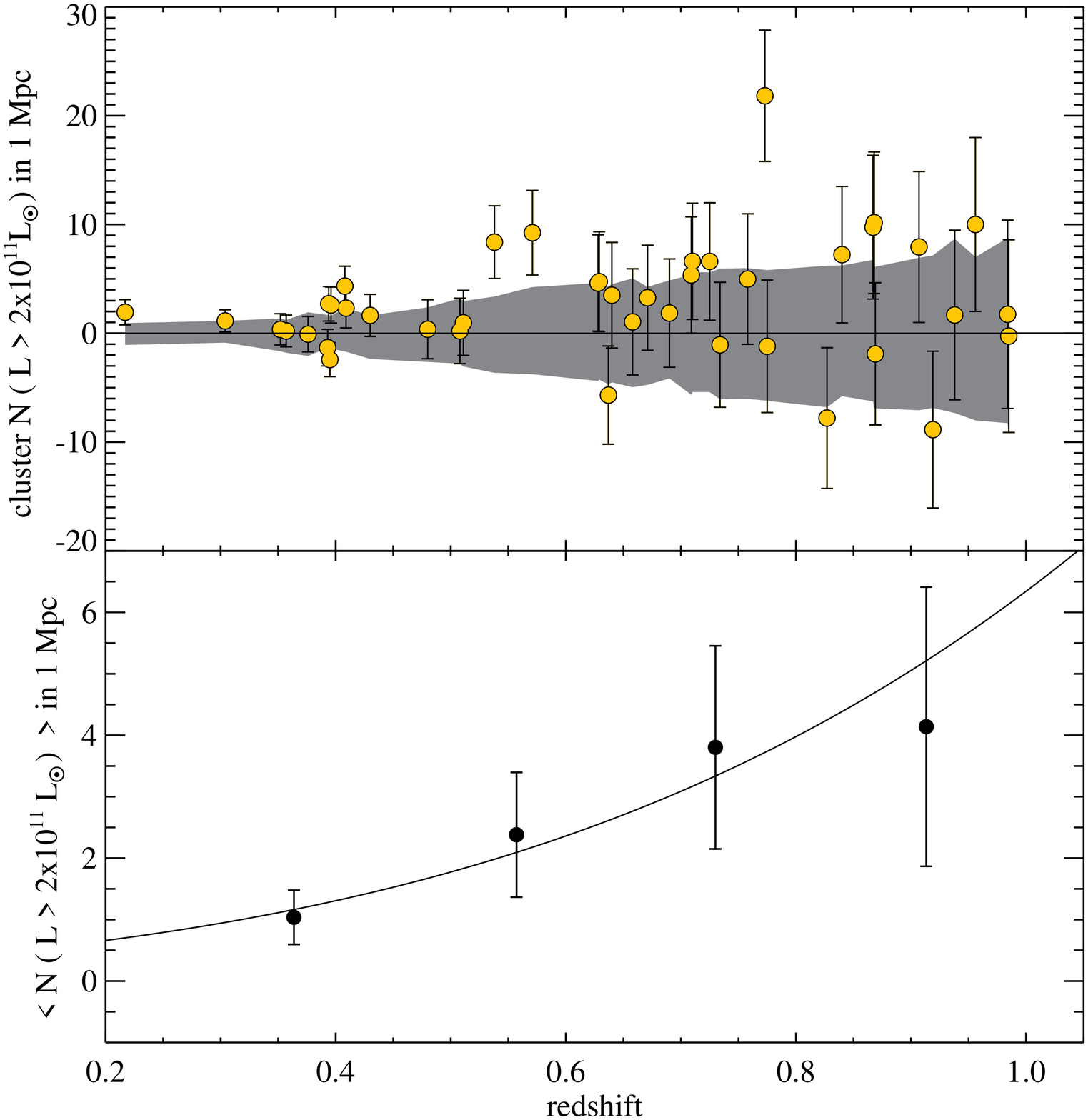}
\caption{The number of 24$\mu$m galaxies above an inferred luminosity of 2$\times$20$^{11}$ L$_\odot$ within the same radius of 1Mpc for all clusters.  This plot is the same as the upper panel in Figure 4, however it does not include the mass normalization and does not alter the counting radius according to the cluster richness.  The solid line is the best-fit power law function of N $\sim$ (1+z)$^{4.4\pm1.9}$.   \label{raw}}
\end{figure}




\clearpage

\end{document}